\documentstyle[jkas]{article}

\beginpage{1}
\endpage{00}
\year{2011}\volume{00}\month{????????}
\runningauthor {J. Ryu \& M.G. Lee} 
\runningtitle{Five Open Clusters in the SDSS}

\month{????????} \year{2011} \volume{00}
\beginpage{1}\endpage{00}
\date{Received ???????? ?, 2011; Accepted ??????? ??, 2011}

\begin{document}
\title{A Photometric Study of Five Open Clusters in the SDSS} 
\author{Jinhyuk Ryu and Myung Gyoon Lee} 
\address{Astronomy Program, Department of Physics and Astronomy, Seoul National University,
Seoul 151-742, Korea\\ {\it E-mail : ryujh@astro.snu.ac.kr and mglee@astro.snu.ac.kr}}

\address{\normalsize{\it (Received ????? ?, 2011; Accepted ????? ??, 2011)}}
\offprints{M.G. Lee}
%--------------------------------------------------------------------
\abstract{We present a photometric study of five open clusters (Czernik 5, Alessi 53, Berkeley 49, Berkeley 84, and Pfleiderer 3) in the Sloan Digital Sky Survey. The position and size of these clusters are determined using the radial number density profiles of the stars, and the member stars of the clusters are selected using the proper motion data in the literature. We estimate the reddening, distance, and age of the clusters based on the isochrone fitting in the color-magnitude diagram. The foreground reddenings for these clusters are estimated to be $E(B-V)=0.71-1.55$ mag. The distances to these clusters are derived to be $2.0-4.4$ kpc, and their distances from the Galactic center range from 7.57 kpc to 12.35 kpc. Their ages are in the range from 250 Myr to 1 Gyr. Berkeley 49 and Berkeley 84 are located in the Orion spur, Czernik 5 is in the Perseus arm, and Pfleiderer 3 and Alessi 53 are at beyond the Perseus arm.}

\maketitle

%--------------------------------------------------------------------
\section{INTRODUCTION}

	Open clusters (OCs) are groups of tens to a few thousand stars with sizes of several parsecs. Stars in an OC are born simultaneously in a giant molecular cloud, which is usually found at the Galactic disk. Therefore these OCs are excellent tracers of the formation and evolution of the Galactic disk. As an example, \citet{pia10} found that the age distribution of the OCs in our Galaxy shows two prominent peaks at 10--15 Myrs and 1.5 Gyrs, providing an evidence of enhanced formation episodes of the OCs at these epochs. There are more than two thousands of OCs and OC candidates listed in the catalog of optically visible open clusters and candidates by Dias, Alessi, Moitinho, and L\'epine 2002 (hereafter DAML02\footnote{Last updated at November 24th, 2010.}). However, according to DAML02, physical parameters such as the reddening, distance, and age of the clusters are still unknown for about half of them. Therefore, it is needed to determine the physical parameters of these OCs for enhancing our knowledge about the Galaxy structure.
	
	To determine the physical parameters of OCs, we can use photometry for individual OCs that are observed by ourselves (\citealt{lee97}; \citealt{pa99}; \citealt{ann02}). However, it is less efficient than using wide field survey data. It contains enormous amount of data which cover a large area of the sky. One of the widely-used surveys is the Two Micron All Sky Survey (2MASS; \citealt{skr06}). There are many studies about OCs based on the 2MASS, including discoveries of new OC candidates (\citealt{kr06}; \citealt{fro07}; \citealt{kop08}), and estimations of physical parameters of OCs (\citealt{ta08}; \citealt{ta09}; \citealt{cam09}).
	
	The Sloan Digital Sky Survey (SDSS; \citealt{yo00}) is another excellent resource for the OC study. The SDSS covers a quarter of the whole sky and contains a few hundred million stars. Nevertheless, there are only a few studies about OCs using the SDSS until now. For examples, \citet{an08} presented fiducial sequences of 17 Galactic globular clusters and 3 well-known OCs (NGC 2420, NGC 6791, and M67), and \citet{an09} made theoretical isochrones and compared them with data in \citet{an08}.

	In this paper, using the SDSS, we present a photometric study of five OCs in our Galaxy those were little studied: Czernik 5, Alessi 53, Berkeley 49, Berkeley 84, and Pfleiderer 3. We determine the physical parameters of these clusters, using the $ugriz$ photometry we derive from the SDSS images. Three of them (Czernik 5, Berkeley 49, and Berkeley 84) were studied previously using 2MASS data (\citealt{ta08}; \citealt{ta09}; \citealt{cam09}), but none of them were studied in the optical band. During this study, \citet{su10} presented BVI photometry of Berkeley 49. No study is available for Alessi 53 and Pfleiderer 3 as yet.
	
	This paper is organized as follows. Section 2 describes data, the selection of target clusters, and how to derive photometry of the target clusters. Section 3 explains the methods to determine the center and size of each cluster, and derive physical parameters of the clusters. In Section 4, we present color-magnitude diagrams (CMD) of these clusters. Then we derive foreground reddening, distance, and age of these clusters from the CMDs. We discuss the properties of the target clusters, and their locations in the Milky Way in Section 5. Finally primary results are summarized in Section 6.

%%%%%%%%%%%%%%%%%%%%%%%%%%%%%%%%%%%%%%%%%%%%%%%%%%%%%%%%%%%%%%%%%%%%%%%%%%%%%%%%%%%
%%%%%%%%%%%%%%%%%%%%%%%%%%%%%%%%%%%%%%%%%%%%%%%%%%%%%%%%%%%%%%%%%%%%%%%%%%%%%%%%%%%
\section{DATA AND DATA REDUCTION}

	We used $u$, $g$, $r$, $i$, and $z$ images in the SDSS for this study. The SDSS obtains images and spectra of sources using a 2.5-m telescope at the Apache Point Observatory. It covers a quarter of the entire sky for survey, and limiting magnitudes of photometry are 22.0, 22.2, 22.2, 21.3, and 20.5 mag for $u$, $g$, $r$, $i$, and $z$ bands, respectively. These magnitudes correspond to 95\% detection repeatability for point sources. The field of view for a chip is 13$\arcmin$.51$\times$8$\arcmin$.98, and the pixel scale is 0\arcsec.396. An image is taken by the drift scan with an integration time of 54 seconds.
	
	We selected the target clusters among 170 OCs in the SDSS Data Release 7 (DR7; \citealt{aba09}) with following criteria: (1) being located far from the edge of the survey area, (2) being distinguished clearly from backgrounds, (3) being poorly studied, (4) not being saturated, and (5) not having too large angular size (smaller than $\sim10\arcmin$). Finally we chose 5 OCs: Czernik 5, Alessi 53, Berkeley 49, Berkeley 84, and Pfleiderer 3.

	The SDSS provides point-spread-function (PSF) fitting magnitudes of point sources. However, standard SDSS photometric pipelines \citep{lup02} do not provide PSF magnitudes of some stars in the crowded region like a central region of a star cluster. Therefore we decided to derive photometry of these clusters by ourselves rather than using the SDSS DR7 catalog. 	
	
	We derived PSF magnitudes from the SDSS images using the IRAF\footnote{IRAF is distributed by NOAO, which are operated by the Association of Universities for Research in Astronomy, Inc. under contract with the National Science Foundation.}/DAOPHOT \citep{st87}. We obtained FITS files of the chosen clusters from the SDSS Data Archive Server (DAS). These images were already bias subtracted, dark corrected, and flat fielded by pipelines. 
	
\begin{figure*}[htp]
	\centering
	\epsfxsize=14cm \epsfbox{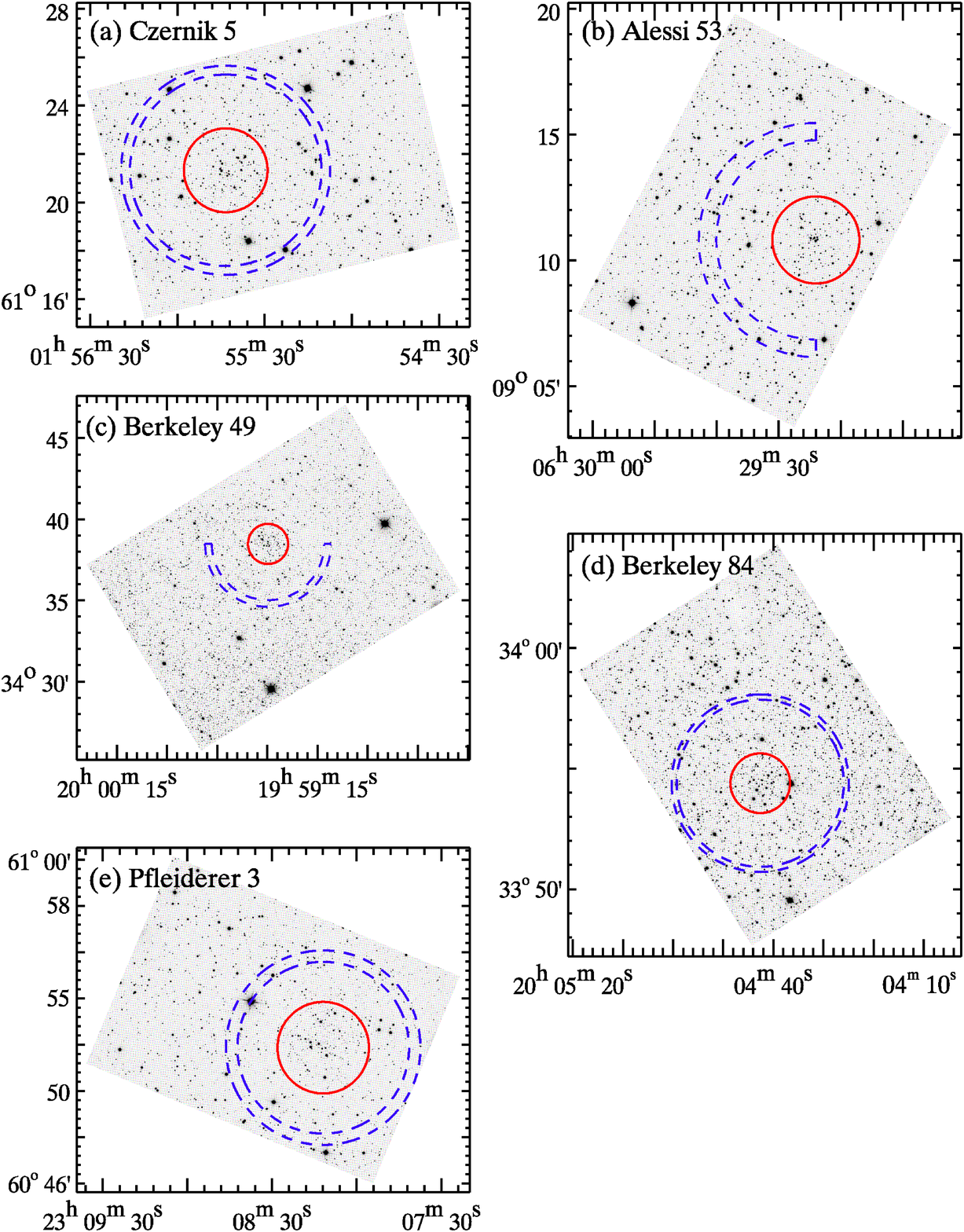}
	\caption{Gray-scale maps of the $r$ band images for target clusters, which are obtained from the SDSS DAS. Images are rotated and combined using SWarp \citep{ber02}. The x-axis and y-axis denote the RA and Dec coordinate, respectively. Each solid circle indicates the size of the cluster, and the region within dashed circles represents a background region which has the same area as that of the cluster region. The name of clusters and the radius of the inner circle or arc in each panel are as follows: (a) Czernik 5 ($R_{in}=4\arcmin$), (b) Alessi 53 ($R_{in}=4\arcmin$), (c) Berkeley 49 ($R_{in}=3\arcmin.5$), (d) Berkeley 84 ($R_{in}=3\arcmin.5$), and (e) Pfleiderer 3 ($R_{in}=3\arcmin.75$). \label{fig:image}}
	\end{figure*}

	Figure \ref{fig:image} shows gray-scale maps of $r$ band images of the clusters. Pfleiderer 3 was observed twice, and we combined the two images. For each field containing the clusters, we derived photometry independently. We used the gain and readout noise of the SDSS chips from Table 4 of \citet{an08}. After deriving the magnitudes in each band, we selected point sources that are detected at least in two bands. 

	We derived zero-point differences using the sources that are $r\leq17$ and common between ours and the SDSS DR7 point source catalogs. Zero-points, rms errors, and the number of stars used for the standard calibration are listed in Table \ref{tab:coeff}. All rms errors of the standard calibration is less than 0.004. Table \ref{tab:sample} lists the $ugriz$ photometry of these five clusters.
	
	\footnotesize
	\begin{deluxetable}{c c c c|c c c|c c c}
	\tablecolumns{10}
	\tablewidth{0pc}
	\tablecaption{Transformation coefficients \label{tab:coeff}}
	\tablehead{
		\colhead{} &		
		\multicolumn{3}{c}{Czernik 5} &
		\multicolumn{3}{c}{Alessi 53} &
		\multicolumn{3}{c}{Berkeley 49 (Field 1)}
	}
	\startdata
	Filter & Z & rms & \# of stars & Z & rms & \# of stars & Z & rms & \# of stars\\
	\hline
	u$\arcmin$ & 2.232 & 0.004 & 96 & 1.966 & 0.002 & 141 & 2.505 & 0.003 & 70\\
	g$\arcmin$ & 1.074 & 0.001 & 258 & 1.160 & 0.002 & 309 & 1.067 & 0.001 & 84\\
	r$\arcmin$ & 1.385 & 0.001 & 444 & 1.374 & 0.000 & 448 & 1.427 & 0.001 & 118\\
	i$\arcmin$ & 1.764 & 0.001 & 618 & 1.697 & 0.001 & 635 & 1.693 & 0.003 & 130\\
	z$\arcmin$ & 3.460 & 0.000 & 626 & 3.001 & 0.001 & 673 & 3.409 & 0.002 & 130\\
	\hline
	\\
	& \multicolumn{3}{c}{Berkeley 49 (Field 2)} & \multicolumn{3}{c}{Berkeley 84} &	\multicolumn{3}{c}{Pfleiderer 3}\\
	\hline
	Filter & Z & rms & \# of stars & Z & rms & \# of stars & Z & rms & \# of stars\\
	\hline
	u$\arcmin$ & 2.495 & 0.003 & 40 & 2.459 & 0.003 & 37 & 1.899 & 0.002 & 105\\
	g$\arcmin$ & 1.069 & 0.002 & 44 & 1.055 & 0.002 & 45 & 1.094 & 0.001 & 219\\
	r$\arcmin$ & 1.406 & 0.002 & 51 & 1.396 & 0.001 & 56 & 1.340 & 0.001 & 329\\
	i$\arcmin$ & 1.637 & 0.003 & 57 & 1.738 & 0.002 & 54 & 1.710 & 0.001 & 427\\
	z$\arcmin$ & 3.413 & 0.004 & 61 & 3.373 & 0.004 & 59 & 3.039 & 0.001 & 477\\
	\enddata
	\end{deluxetable}	
	\normalsize
	
	\scriptsize
	\begin{deluxetable}{c c c c c c c c c c c c c c}
	\tablecolumns{14}
	\tablewidth{0pc}
	\tablecaption{$ugriz$ Photometry of five open clusters\tablenotemark{a} \label{tab:sample}}
	\tablehead{
		\colhead{ID} &		
		\colhead{RA(J2000)} &
		\colhead{DEC(J2000)} &
		\colhead{$u$} &
		\colhead{$\epsilon_u$} &
		\colhead{$g$} &
		\colhead{$\epsilon_g$} &
		\colhead{$r$} &
		\colhead{$\epsilon_r$} &
		\colhead{$i$} &
		\colhead{$\epsilon_i$} &
		\colhead{$z$} &
		\colhead{$\epsilon_z$} &
		\colhead{Membership\tablenotemark{b}}
	}
	\startdata
		\\		
		\multicolumn{14}{l}{\footnotesize\textbf{(a) Czernik 5}}\\
		\hline
		\scriptsize
		7 & 01:54:25.22 & +61:19:46.2 & 20.011 & 0.041 & 18.236 & 0.005 & 17.200 & 0.005 & 17.498 & 0.111 & --- & --- & 1\\
		185 & 01:54:35.82 & +61:20:06.3 & 18.864 & 0.015 & 16.744 & 0.003 & 15.667 & 0.004 & 15.167 & 0.003 & 14.882 & 0.004 & 1\\
		326 & 01:54:40.78 & +61:25:14.2 & 18.463 & 0.011 & 16.423 & 0.002 & 15.701 & 0.004 & 15.315 & 0.002 & 15.024 & 0.004 & 1\\
		421 & 01:54:43.31 & +61:19:26.3 & 19.347 & 0.019 & 17.173 & 0.004 & 16.212 & 0.004 & 15.693 & 0.007 & 15.230 & 0.004 & 1\\
		440 & 01:54:43.76 & +61:18:46.1 & 17.557 & 0.008 & 15.944 & 0.003 & 15.086 & 0.004 & 14.665 & 0.004 & 14.434 & 0.003 & 1\\
		\hline
		\\		
		\multicolumn{14}{l}{\footnotesize\textbf{(b) Alessi 53}}\\
		\hline
		\scriptsize
		2 & 06:29:03.20 & +09:15:26.0 & 18.764 & 0.022 & 17.228 & 0.007 & 16.504 & 0.014 & 16.084 & 0.003 & 15.850 & 0.008 & 1\\
		129 & 06:29:08.27 & +09:14:00.6 & 21.990 & 0.165 & 19.872 & 0.011 & 18.901 & 0.013 & 18.297 & 0.005 & 17.925 & 0.019 & 1\\
		217 & 06:29:10.81 & +09:12:16.3 & --- & --- & 21.939 & 0.062 & 20.615 & 0.042 & 19.825 & 0.020 & 19.463 & 0.056 & 2\\
		262 & 06:29:11.74 & +09:13:18.9 & 19.821 & 0.032 & 18.043 & 0.004 & 17.235 & 0.007 & 16.876 & 0.012 & 16.625 & 0.009 & 1\\
		350 & 06:29:13.48 & +09:13:43.4 & 18.472 & 0.014 & 16.939 & 0.003 & 16.252 & 0.006 & 15.958 & 0.009 & 15.743 & 0.006 & 1\\
		\hline
		\\		
		\multicolumn{14}{l}{\footnotesize\textbf{(c) Berkeley 49}}\\
		\hline
		\scriptsize
		114 & 19:58:36.92 & +34:36:30.3 & 21.338 & 0.190 & 18.870 & 0.007 & 17.724 & 0.008 & 17.124 & 0.012 & 16.611 & 0.010 & 1\\
		262 & 19:58:39.12 & +34:36:10.3 & 18.513 & 0.022 & 16.573 & 0.003 & 15.602 & 0.006 & 15.162 & 0.008 & 14.873 & 0.005 & 1\\
		692 & 19:58:43.06 & +34:37:09.3 & --- & --- & 22.495 & 0.070 & 20.740 & 0.033 & 19.974 & 0.030 & 19.249 & 0.040 & 1\\
		810 & 19:58:43.94 & +34:38:52.2 & 17.551 & 0.015 & 15.771 & 0.003 & 14.886 & 0.006 & 14.492 & 0.010 & 14.262 & 0.003 & 1\\
		1006 & 19:58:45.45 & +34:33:57.5 & --- & --- & 21.399 & 0.029 & 19.680 & 0.014 & 18.776 & 0.012 & 18.104 & 0.016 & 1\\
		\hline
		\\		
		\multicolumn{14}{l}{\footnotesize\textbf{(d) Berkeley 84}}\\
		\hline
		\scriptsize
		48 & 20:04:06.70 & +33:53:18.6 & 19.770 & 0.033 & 18.008 & 0.006 & 17.114 & 0.004 & 16.533 & 0.006 & 16.316 & 0.008 & 1\\
		104 & 20:04:07.70 & +33:53:13.6 & 17.634 & 0.011 & 16.066 & 0.002 & 15.049 & 0.004 & 14.356 & 0.005 & 14.003 & 0.006 & 1\\
		236 & 20:04:09.33 & +33:52:46.1 & 16.651 & 0.007 & 15.400 & 0.004 & 14.993 & 0.003 & 14.698 & 0.003 & 14.725 & 0.006 & 1\\
		331 & 20:04:10.49 & +33:54:43.3 & 18.273 & 0.014 & 16.776 & 0.003 & 16.174 & 0.006 & 15.816 & 0.005 & 15.767 & 0.007 & 1\\
		509 & 20:04:11.92 & +33:52:28.4 & 18.187 & 0.011 & 16.401 & 0.019 & 16.002 & 0.006 & 15.702 & 0.015 & 15.663 & 0.016 & 1\\		
		\hline	
		\\		
		\multicolumn{14}{l}{\footnotesize\textbf{(e) Pfleiderer 3}}\\
		\hline	
		\scriptsize
		72 & 23:07:32.16 & +60:52:32.3 & --- & --- & 21.408 & 0.046 & 19.630 & 0.020 & 18.785 & 0.012 & 18.240 & 0.014 & 2\\
		108 & 23:07:35.17 & +60:55:21.9 & 19.516 & 0.025 & 17.402 & 0.006 & 16.276 & 0.006 & 15.690 & 0.007 & 15.275 & 0.011 & 1\\
		216 & 23:07:40.73 & +60:53:33.9 & 19.588 & 0.021 & 16.382 & 0.004 & 14.788 & 0.004 & 14.030 & 0.004 & 13.473 & 0.003 & 1\\
		285 & 23:07:42.75 & +60:49:28.7 & 17.736 & 0.007 & 16.045 & 0.009 & 15.356 & 0.003 & 15.072 & 0.003 & 14.811 & 0.006 & 1\\
		372 & 23:07:45.70 & +60:48:35.5 & 21.406 & 0.090 & 19.337 & 0.007 & 18.014 & 0.007 & 17.385 & 0.008 & 16.962 & 0.010 & 1\\
	\enddata
	\tablenotetext{a}{This is a sample of the full tables, which are compiled from the catalog of each cluster. The full tables will be available from the authors upon request.}
	\tablenotetext{b}{1 and 2 represent cluster members and non-members, respectively.}	
	\end{deluxetable}
	\normalsize

%%%%%%%%%%%%%%%%%%%%%%%%%%%%%%%%%%%%%%%%%%%%%%%%%%%%%%%%%%%%%%%%%%%%%%%%%%%%%%%%%%%
%%%%%%%%%%%%%%%%%%%%%%%%%%%%%%%%%%%%%%%%%%%%%%%%%%%%%%%%%%%%%%%%%%%%%%%%%%%%%%%%%%%
\section{METHOD}

\subsection{Membership Determination}

\begin{figure*}[htp]
	\centering
	\epsfxsize=14cm	\epsfbox{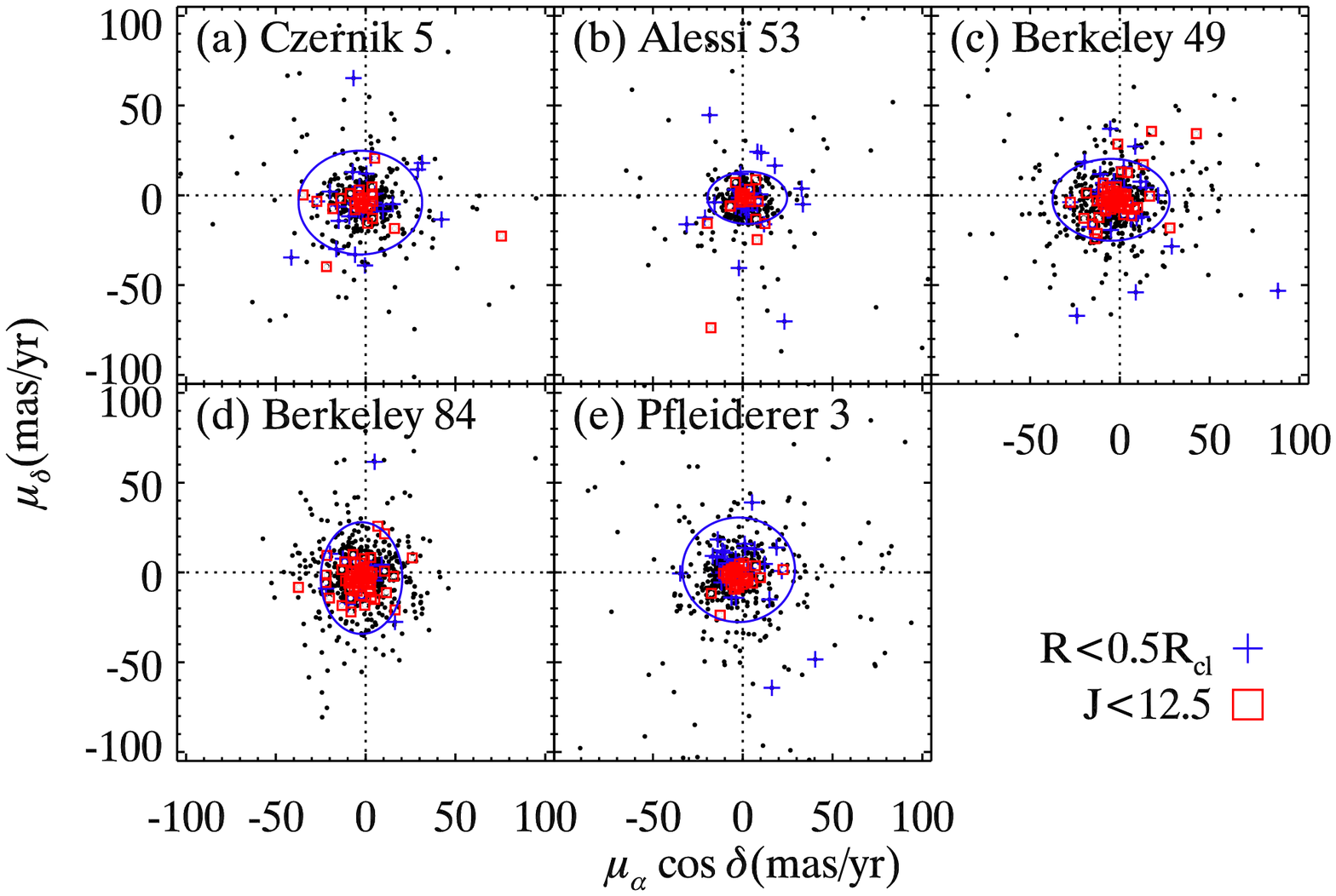}
	\caption{Vector point diagrams of the proper motions in the field of each cluster: (a) Czernik 5, (b) Alessi 53, (c) Berkeley 49, (d) Berkeley 84, and (e) Pfleiderer 3. The ellipses around the zero point are the boundary to exclude non-member stars. Dots represent the stars with $\textrm{J}<14$, squares represent the bright stars with $\textrm{J}<12.5$, and plus symbols represent the stars inside the half radius of the cluster. These symbols indicate that they are more probable members than other stars of the cluster, but some of them are lying the outside of the ellipse. \label{fig:pm}}
	\end{figure*}

	We selected member stars of the clusters using proper motion data in the PPMXL \citep{roe10b}. The PPMXL is an astrometric catalog prepared by combining USNO-B1.0 and the 2MASS astrometry. The catalog provides positions and absolute proper motions. It is complete down to about $V=20$ mag. The average and dispersion of the proper motions were estimated using the stars in the field of each cluster with following conditions: $\textrm{J}<14$, $R\leq2R_{\mathrm{cl}}$, where $R_{\mathrm{cl}}$ represents the radius of the cluster, and $-50\leq\left[ \mu_{\alpha}\cos{\delta}\textrm{, }\mu_{\delta}\right]\leq50$. 
	Based on these values, we defined the boundary of the proper motion. Figure \ref{fig:pm} shows vector point diagrams of the proper motion for the five clusters. Most stars in each cluster are distributed near the zero point (0,0), but the average values of the proper motions are slightly off from the zero point. Some stars located far from this group may not be members of the cluster. We use an ellipse for defining the boundary, because dispersions along the RA-direction and the Dec-direction are not the same. This ellipse has an average of proper motions as a center, and 3 times of dispersions as axis lengths.
			
\subsection{Determining the center and size of the clusters}
	
\begin{figure*}[htp]
	\centering \epsfxsize=14cm
	\epsfbox{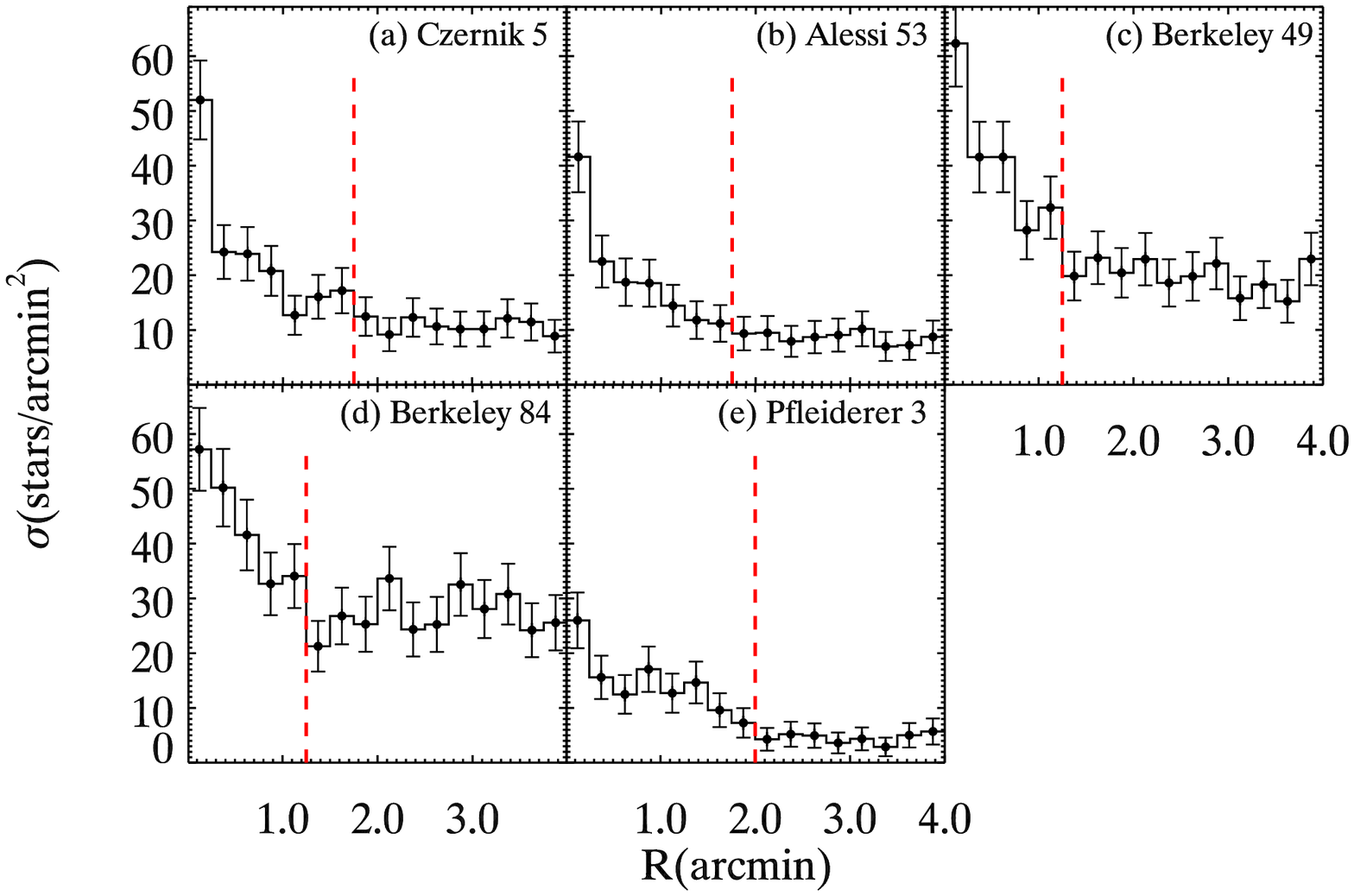}
	\caption{Radial number density profiles for the stars in the clusters: (a) Czernik 5, (b) Alessi 53, (c) Berkeley 49, (d) Berkeley 84, and (e) Pfleiderer 3. The dashed lines represent the size of each cluster. Error bars denote Poisson errors. \label{fig:rdp}}
	\end{figure*}

	We determined the center of the clusters by finding the maximum surface number density of the member stars in each cluster. We investigated the peak value of the number density using the stars within $14\;{\leq}\;r\;{\leq}\;20$ for the clusters except for Pfleiderer 3. The stars in Pfleiderer 3 are fainter than those in other clusters so that we used stars within $18\:{\leq}\:r\:{\leq}\:20$. A searching bin used for finding the maximum is a circle of $R=0\arcmin.25$ radius. If there are more than one local maximum of the number density, we moved the center of clusters to the position of a peak value which is confirmed visually. Using these new centers, we derived radial number density profiles of the clusters, as shown in Figure \ref{fig:rdp}. We determined visually the size of the clusters as where the excess of the number density is indistinguishable from the outer region. The coordinates and sizes of the clusters from DAML02 and from this study are listed in Table \ref{tab:position}. 

	\small
	\begin{deluxetable}{c c c c|c c c|r r l}
	\tablecolumns{10}
	\tablewidth{0pc}
	\tablecaption{The coordinates and sizes of the target clusters \label{tab:position}}
	\tablehead{
		\colhead{} & \multicolumn{3}{c}{DAML02} & \multicolumn{3}{c}{This study} &
		\multicolumn{3}{c}{(This study) -- (DAML02)}\\
		\hline
		\colhead{Name} &
		\colhead{RA} &
		\colhead{DEC} &
		\colhead{$R_{\mathrm{cl}}$} &
		\colhead{RA} &
		\colhead{DEC} &
		\colhead{$R_{\mathrm{cl}}$} & $\Delta$ RA & $\Delta$ DEC & $\Delta R_{\mathrm{cl}}$\\
		\colhead{} &
		\colhead{(h m s)} &
		\colhead{($\arcdeg$ $\arcmin$ $\arcsec$)} &
		\colhead{($\arcmin$)} &
		\colhead{(h m s)} &
		\colhead{($\arcdeg$ $\arcmin$ $\arcsec$)} &
		\colhead{($\arcmin$)} & \colhead{} & \colhead{} & \colhead{}
	}
	\startdata
		Czernik 5 & 01 55 06 & +61 20 00 & 0.75 & 01 55 43.4 & +61 21 21 & 1.75 & $561\arcsec.0$ & $81\arcsec$ & $\;\;\:1\arcmin.00$\\
		Alessi 53 & 06 29 24 & +09 10 55 & 1.4 & 06 29 24.0 & +09 10 49 & 1.75 &   $0\arcsec.0$ & $-6\arcsec$ & $\;\;\:0\arcmin.35$\\
		Berkeley 49 & 19 59 31 & +34 38 48 & 2.0 & 19 59 29.8 & +34 38 30 & 1.25 & $-18\arcsec.0$ & $-18\arcsec$ & $-0\arcmin.75$\\
		Berkeley 84 & 20 04 43 & +33 54 18 & 4.5 & 20 04 42.6 & +33 54 24 & 1.25 &  $-6\arcsec.0$ & $6\arcsec$ & $-3\arcmin.25$\\
		Pfleiderer 3 & 23 08 11 & +60 52 24 & 2.0 & 23 08 11.8 & +60 51 54 & 2.00 &  $12\arcsec.0$ & $-30\arcsec$ & $\;\;\:0\arcmin.00$\\
	\enddata
	\end{deluxetable}
	\normalsize

\subsection{Determining the physical parameters}

	We derived the physical parameters of these clusters using the following methods. We determined the foreground reddening using the Padova isochrones \citep{gi04} in the CMD and the Zero Age Main Sequence (ZAMS) in the Color-Color Diagram (CCD). The extinction law in \citet{an08} is used for determining reddenings: ${A_u}/{A_V}=1.574$, ${A_g}/{A_V}=1.189$, ${A_r}/{A_V}=0.877$, ${A_i}/{A_V}=0.673$, and ${A_z}/{A_V}=0.489$,  assuming ${R_V}\equiv{A_V}/E(B-V)=3.1$. After determining the foreground reddening, we estimated the distance and age for the clusters using the isochrone fitting in the CMD. For this fitting, we need to know the metallicity of the clusters. However, it is not easy to derive a metallicity from photometry. Therefore we adopted the metallicity for the clusters using the relation between [Fe/H] and the galactocentric distance $R_{\mathrm{GC}}$. 
	
\begin{figure}[tp]
	\centering \epsfxsize=8cm
	\epsfbox{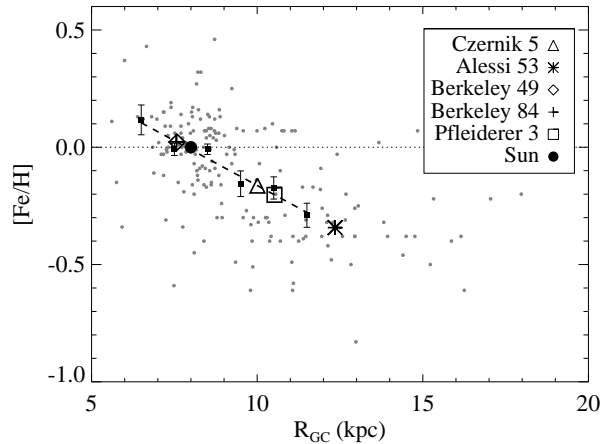}
	\caption{Metallicity vs. Galactocentric distance diagram for 186 OCs in DAML02. Filled squares and error bars represent the mean metallicities. Large symbols represent the assumed metallicity of the target clusters in this study. \label{fig:gradient}}
	\end{figure}

	Figure \ref{fig:gradient} displays [Fe/H] and $R_{\mathrm{GC}}$ relation based on data for 186 clusters with [Fe/H] values in DAML02. We derived the mean metallicity for each 1 kpc-size bin as a function of $R_{\mathrm{GC}}$, as plotted in the same figure. The mean metallicity decreases as $R_{\mathrm{GC}}$ increases for
$R_{\mathrm{GC}}\lesssim12$ kpc, and appears to be constant at [Fe/H] $\approx -0.3$ thereafter. We fitted the data for $R_{\mathrm{GC}}\lesssim12$ kpc with a linear function, obtaining $<$[Fe/H]$>=(-0.076\pm0.013) R_{\mathrm{GC}} + (0.600\pm0.116)$ with rms = 0.029. This value for the radial gradient is similar to previous estimates (\citealt{fri02}; \citealt{che03}; \citealt{and11}). We adopt the metallicity for each cluster using this relation: [Fe/H] = +0.022, +0.023, -0.164, -0.203, and -0.344 (Z=0.019, 0.019, 0.016, 0.016, 0.013) for Berkeley 49, Berkeley 84, Czernik 5, Pleiderer 3, and Alessi 53, respectively.
		
%%%%%%%%%%%%%%%%%%%%%%%%%%%%%%%%%%%%%%%%%%%%%%%%%%%%%%%%%%%%%%%%%%%%%%%%%%%%%%%%%%%
%%%%%%%%%%%%%%%%%%%%%%%%%%%%%%%%%%%%%%%%%%%%%%%%%%%%%%%%%%%%%%%%%%%%%%%%%%%%%%%%%%%
\section{RESULTS}

	We describe the results we derived applying the above methods for each cluster in the following.

\subsection{Czernik 5}
	
	The coordinates for the center of this Czernik 5 we derived are ($\alpha$ = 01h 55m 43.4s, $\delta$ = +61$\arcdeg$ 21$\arcmin$ 21$\arcsec$), which are different by 561$\arcsec$.0 in RA and 81$\arcsec$ in Dec from the values in DAML02. The radial number density profile of Czernik 5 derived using the new cluster center is shown in Figure \ref{fig:rdp}(a). From this figure, the radius of Czernik 5 is determined to be $R_{\mathrm{cl}}=1\arcmin.75\pm0.13$.
	
\begin{figure}[tp]
	\centering \epsfxsize=8cm
	\epsfbox{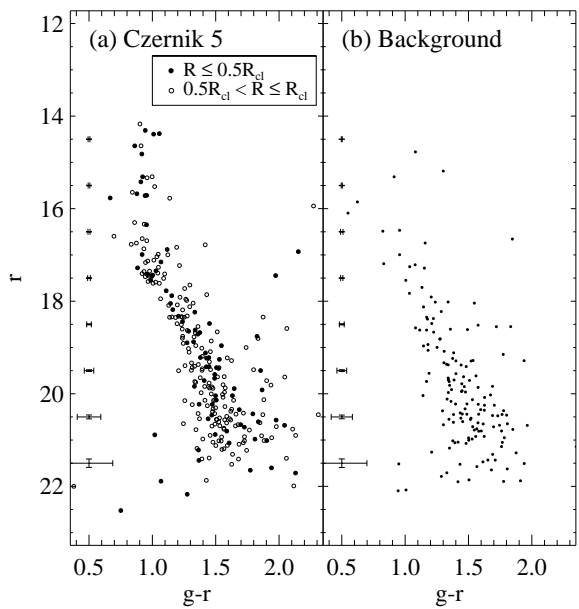}
	\caption{(a) $r-(g-r)$ CMD of the stars in Czernik 5. Open circles are stars inside the radius of Czernik 5, and filled dots are those inside the half radius of the cluster. (b) $r-(g-r)$ CMD of the stars in the background region, as shown in Figure \ref{fig:image}(a). Mean errors are marked at the left side of each panel. \label{fig:cz5cmd}}
	\end{figure}

	Figure \ref{fig:cz5cmd} shows (a) an $r-(g-r)$ CMD of Czernik 5, and (b) a CMD of the background region which has the same area as that of the cluster region, as shown in Figure \ref{fig:image}(a). Main sequences are found on both CMDs, but more stars are seen in the cluster's CMD. It is notable that the cluster CMD shows about two dozen bright stars at $0.8\leq (g-r)\leq 1.2$ and $14\leq r \leq17$, while the background CMD shows few. Most of these stars are considered to be the members of the cluster. About half of these bright stars are located close to the cluster center, within the half of the radius.

\begin{figure}[htp]
	\centering \epsfxsize=8cm
	\epsfbox{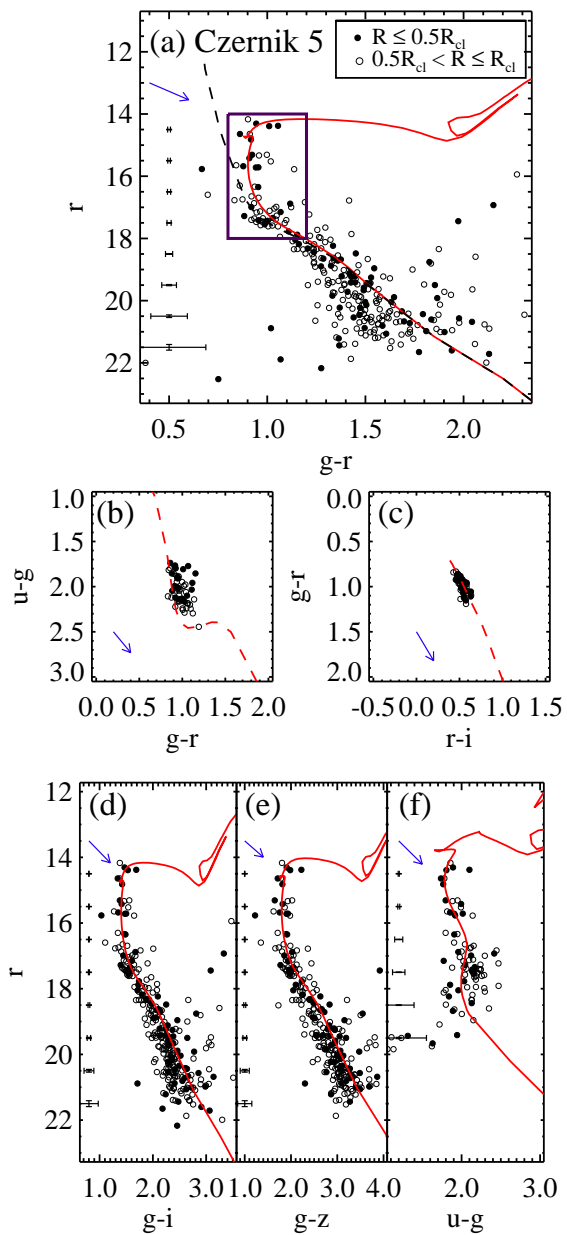}
	\caption{CMDs and CCDs of Czernik 5. Symbols represent the same as in Figure \ref{fig:cz5cmd}. The solid line represents an isochrone for log(age[yr])=8.45 and $Z=0.016$, and the dashed line represents the ZAMS. The arrow in each panel represents a reddening vector. (a) Isochrone fitting for Czernik 5 in the $r-(g-r)$ CMD. The isochrone is shifted according to $E(B-V)=1.20$ and $(m-M)_0=12.2$. The solid box contains main sequence stars that are plotted in the following CCDs. (b) $(u-g)-(g-r)$ CCD. (c) $(g-r)-(r-i)$ CCD. (d) $r-(g-i)$ CMD. (e) $r-(g-z)$ CMD. (f) $r-(u-g)$ CMD.\label{fig:cz5}}
	\end{figure}

	Figure \ref{fig:cz5} shows the result of the isochrone fitting. We derived its age, log(age[yr])=8.45, and the reddenings: $E(B-V)=1.20 \pm 0.05$, $1.20 \pm 0.05$, $1.15 \pm 0.05$, and $0.85 \pm 0.05$ from $r-(g-r)$ CMD, $r-(g-i)$ CMD, $r-(g-z)$ CMD, and $r-(u-g)$ CMD, respectively. Corresponding distance moduli we derived are all $(m-M)_0=12.2 \pm 0.2$. These values are listed in Table \ref{tab:cz5}. The reddening values derived from the first two CMDs agree, while the latter two have somewhat smaller values. This trend is similarly seen in other clusters. This indicates that the theoretical colors in the Padova isochrones, $(g-z)$ and $(u-g)$ are too red by $\sim 0.1$. Hereafter we adopt the values for the reddening and distance modulus derived from $r-(g-r)$ CMD for further analysis, because this CMD is considered to be most reliable for cluster analysis \citep{an08}. Figure \ref{fig:cz5}(b) and (c) display the $(u-g)-(g-r)$ CCD and $(g-r)-(r-i)$ CCD for the bright main sequence stars located within the solid box in Figure \ref{fig:cz5}(a). They are well fitted by the ZAMS shifted according to the reddening $E(B-V)=1.20$. 
	
	\small
	\begin{table}[bp]
	\begin{center}
	\centering
	\caption{Estimates for reddenings and distance moduli of Czernik 5 based on different colors \label{tab:cz5}}
	\doublerulesep2.0pt
	\renewcommand\arraystretch{1.5}
	\begin{tabular}{c c c}
	\hline \hline
	CMD & $E(B-V)$ & $(m-M)_0$\\
	\hline
	$r-(g-r)$ & 1.20$\pm$0.05 & 12.2$\pm$0.2\\
	$r-(g-i)$ & 1.20$\pm$0.05 & 12.2$\pm$0.2\\
	$r-(g-z)$ & 1.15$\pm$0.05 & 12.2$\pm$0.2\\
	$r-(u-g)$ & 0.85$\pm$0.05 & 12.2$\pm$0.2\\
	\hline
	\end{tabular}
	\end{center}
	\end{table}	
	\normalsize

\subsection{Alessi 53}

	This cluster is not easy to discern in the gray-scale map of $r$ band image (Figure \ref{fig:image}(b)). However, the radial number density profile of the stars around this cluster shows clearly an excess in the cluster region. The coordinates for the center of this cluster ($\alpha$ = 06h 29m 24.0s, $\delta$ = +09$\arcdeg$ 10$\arcmin$ 49$\arcsec$) are different by -6$\arcsec$ in Dec from the value in DAML02. Figure \ref{fig:rdp}(b) displays the radial number density profile of this cluster we derived. We determined the radius of Alessi 53 to be $R_{\mathrm{cl}}=1\arcmin.75\pm0.13$.
	
\begin{figure}[tp]
	\centering \epsfxsize=8cm
	\epsfbox{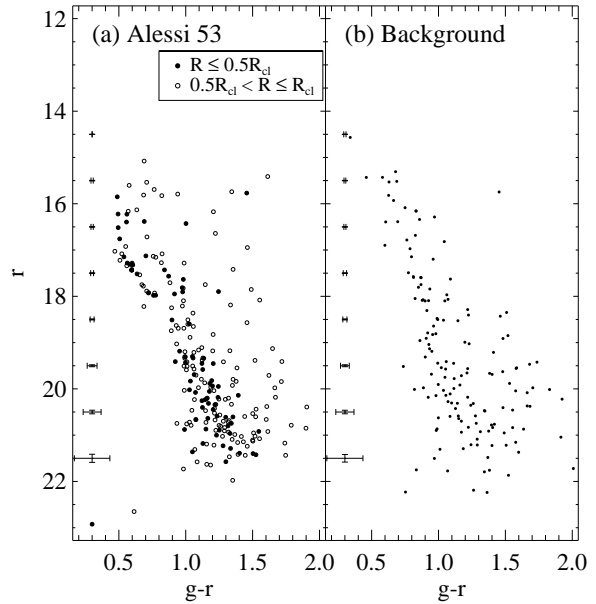}
	\caption{(a) $r-(g-r)$ CMD of the stars in Alessi 53. Open circles are stars inside the radius of Alessi 53, and filled dots are those inside the half radius of the cluster. (b) $r-(g-r)$ CMD of stars in the background region which is shown in Figure \ref{fig:image}(b). Mean errors are marked at the left side of each panel. \label{fig:alessi53cmd}}
	\end{figure}

	The CMD of Alessi 53 in Figure \ref{fig:alessi53cmd} shows a narrow main sequence at $16\leq r \leq18.5$   which is not seen in the CMD of background stars. There is seen a hint of another sequence that is located $\sim0.8$ mag above the main sequence. This may be a binary sequence of this cluster.

\begin{figure}[tp]
	\centering \epsfxsize=8cm
	\epsfbox{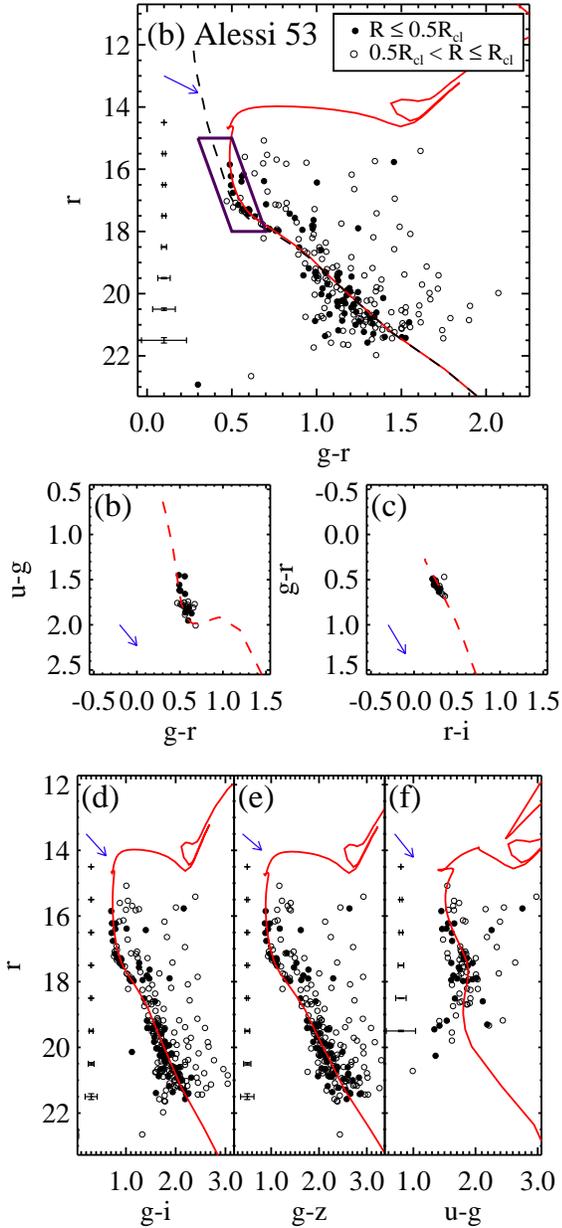}
	\caption{CMDs and CCDs of Alessi 53. Symbols represent the same as in Figure \ref{fig:alessi53cmd}. The solid line represents an isochrone for log(age[yr])=8.4 and $Z=0.013$, and the dashed line represents the ZAMS. The arrow in each panel represents a reddening vector. (a) Isochrone fitting for Alessi 53. The isochrone is shifted according to $E(B-V)=0.80$ and $(m-M)_0=13.3$. The inclined solid box contains main sequence stars which are plotted in the following CCDs. (b) $(u-g)-(g-r)$ CCD. (c) $(g-r)-(r-i)$ CCD. (d) $r-(g-i)$ CMD. (e) $r-(g-z)$ CMD. (f) $r-(u-g)$ CMD.\label{fig:al53}}
	\end{figure}

	Figure \ref{fig:al53} shows the result of the isochrone fitting. We derived its age, log(age[yr])=8.4, and the reddenings: $E(B-V)=0.80 \pm 0.05$, $0.80 \pm 0.05$, $0.75 \pm 0.05$, and $0.70 \pm 0.05$ from $r-(g-r)$ CMD, $r-(g-i)$ CMD, $r-(g-z)$ CMD, and $r-(u-g)$ CMD, respectively. Corresponding distance moduli we derived are $(m-M)_0=13.3 \pm 0.15$ for $r-(g-r)$ CMD and $r-(g-i)$ CMD, and $(m-M)_0=13.5 \pm 0.15$ for $r-(g-z)$ CMD and $r-(u-g)$ CMD. These values are listed in Table \ref{tab:alessi53}. The reddening values derived from the first two CMDs agree, while the latter two have somewhat smaller values. Figure \ref{fig:al53}(b) and (c) display the $(u-g)-(g-r)$ CCD and $(g-r)-(r-i)$ CCD for the bright main sequence stars located within the solid box in Figure \ref{fig:al53}(a). They are well fitted by the ZAMS shifted according to the reddening $E(B-V)=0.80$. 
	
	\small
	\begin{table}[bp]
	\begin{center}
	\centering
	\caption{Estimates for reddenings and distance moduli of Alessi 53 based on different colors \label{tab:alessi53}}
	\doublerulesep2.0pt
	\renewcommand\arraystretch{1.5}
	\begin{tabular}{c c c}
	\hline \hline
	CMD & $E(B-V)$ & $(m-M)_0$\\
	\hline
	$r-(g-r)$ & 0.80$\pm$0.05 & 13.3$\pm$0.15\\
	$r-(g-i)$ & 0.80$\pm$0.05 & 13.3$\pm$0.15\\
	$r-(g-z)$ & 0.75$\pm$0.05 & 13.5$\pm$0.15\\
	$r-(u-g)$ & 0.70$\pm$0.05 & 13.5$\pm$0.15\\
	\hline
	\end{tabular}
	\end{center}
	\end{table}
	\normalsize

\subsection{Berkeley 49}

	The coordinates for the center of Berkeley 49 ($\alpha$ = 19h 59m 29.8s, $\delta$ = +34$\arcdeg$ 38$\arcmin$ 30$\arcsec$) derived in this study are different by -18$\arcsec$.0 in RA and -18$\arcsec$ in Dec from the values in DAML02. From the radial number density profile in Figure \ref{fig:rdp}(c), we determined the radius of Berkeley 49 to be $R_{\mathrm{cl}}=1\arcmin.25\pm0.13$.
	
\begin{figure}[tp]
	\centering \epsfxsize=8cm
	\epsfbox{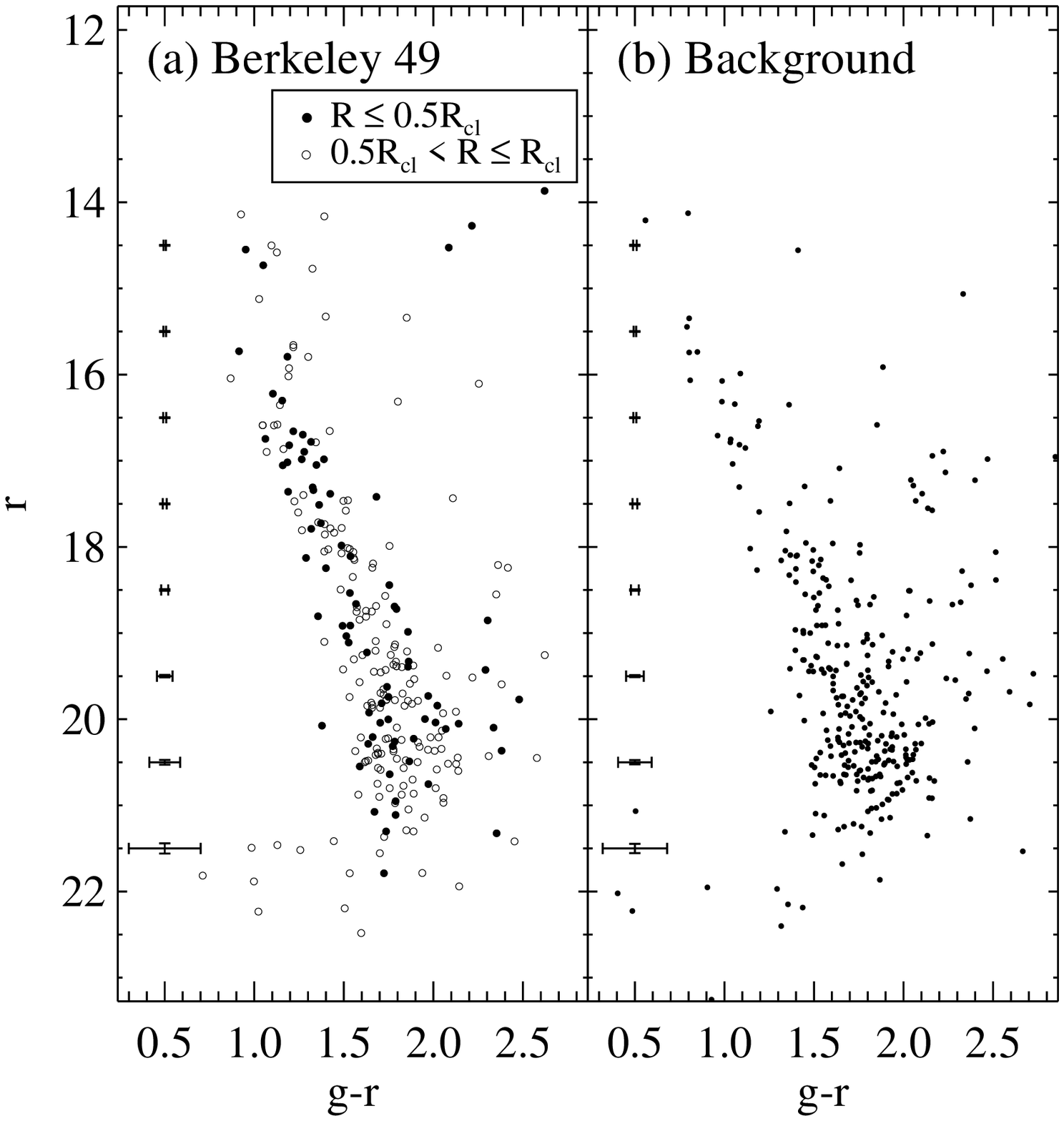}
	\caption{(a) $r-(g-r)$ CMD of the stars in Berkeley 49. Open circles are stars inside the radius of Berkeley 49, and filled dots are stars inside the half radius of the cluster. (b) $r-(g-r)$ CMD of stars in the background region, which is shown in Figure \ref{fig:image}(c). Mean errors are marked at the left side of each panel. \label{fig:be49cmd}}
	\end{figure}

	The CMD of this cluster in Figure \ref{fig:be49cmd} shows a rather broad main sequence. This may be due to severe foreground reddening. The bright main sequence in the cluster CMD is not clearly seen in the background CMD so that most of the stars in this sequence are probably the cluster members. The main sequence turnoff point appears to be located at $(g-r) \sim 1.1$ and $r \sim 16.5$.
		
\begin{figure}[tp]
	\centering \epsfxsize=8cm
	\epsfbox{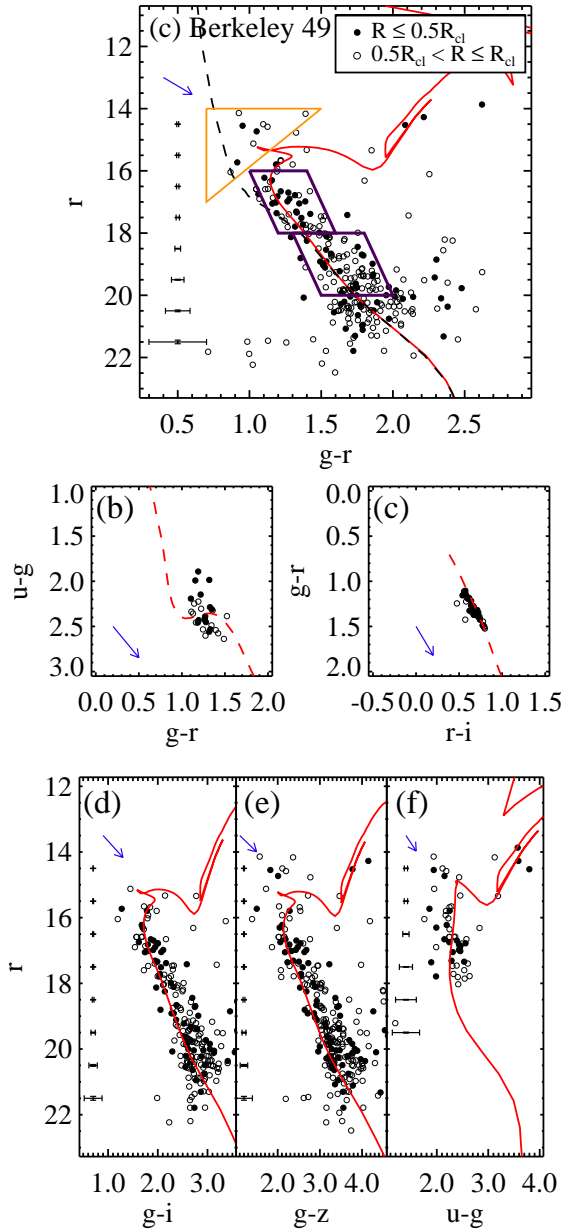}
	\caption{CMDs and CCDs of Berkeley 49. Symbols represent the same as in Figure \ref{fig:be49cmd}. The solid line represents an isochrone for log(age[yr])=8.9 and $Z=0.019$, and the dashed line represents the ZAMS. The arrow in each panel represents a reddening vector. (a) The isochrone fitting for Berkeley 49. The isochrone is shifted according to $E(B-V)=1.18$ and $(m-M)_0=11.6$. The stars located within the triangle are probably the field stars. A parallelogram located within $16<r<18$ contains main sequence stars those are plotted in following CCDs. Another parallelogram, located within $18<r<20$, contains faint main sequence stars. Radial distributions of stars located within each region are discussed in Section 5.3. (b) $(u-g)-(g-r)$ CCD. (c) $(g-r)-(r-i)$ CCD. (d) $r-(g-i)$ CMD. (e) $r-(g-z)$ CMD. (f) $r-(u-g)$ CMD. \label{fig:be49}}
	\end{figure}

	Figure \ref{fig:be49} shows the result of the isochrone fitting. We derived its age, log(age[yr])=8.9, and derived the reddenings and distance moduli from CMDs with different colors: $E(B-V)=1.18 \pm 0.05$ and $(m-M)_0=11.6 \pm 0.2$ from $r-(g-r)$ CMD, $E(B-V)=1.15 \pm 0.05$ and $(m-M)_0=11.6 \pm 0.2$ from $r-(g-i)$ CMD, $E(B-V)=1.10 \pm 0.05$ and $(m-M)_0=11.8 \pm 0.2$ from $r-(g-z)$ CMD, and $E(B-V)=1.05 \pm 0.05$ and $(m-M)_0=11.6 \pm 0.2$ from $r-(u-g)$ CMD. These values are also listed in Table \ref{tab:be49}. The reddening values derived from the CMDs are decreasing from the $r-(g-r)$ CMD to the $r-(u-g)$ CMD. Figure \ref{fig:be49}(b) and (c) display the $(u-g)-(g-r)$ CCD and $(g-r)-(r-i)$ CCD for the bright main sequence stars located within the upper solid box located within $16<r<18$ in Figure \ref{fig:be49}(a). They are well fitted by the ZAMS shifted according to the reddening $E(B-V)=1.18$.
	%The locus in the $(u-g)-(g-r)$ CCD, as shown in Figure \ref{fig:be49}(b), is rather scattered but the ZAMS is fitted to the locus. In the $(g-r)-(r-i)$ CCD, as shown in Figure \ref{fig:be49}(c), the ZAMS is fitted to the locus of Berkeley 49 more tightly than the case of the $(u-g)-(g-r)$ CCD. Figure \ref{fig:be49}(d), (e), and (f) show an isochrone fitting with other color indices. We use slightly different reddenings and distance moduli for fitting an isochrone in each CMD, where $E(B-V)=1.15$ and $(m-M)_0=11.6$ for the $r-(g-i)$ CMD, $E(B-V)=1.1$ and $(m-M)_0=11.8$ for the $r-(g-z)$ CMD, and $E(B-V)=1.05$ and $(m-M)_0=11.6$ for the $r-(u-g)$ CMD. %Note that an isochrone fitting in $r-(u-g)$ CMD is poor.
	% Differences in the reddening may be caused by uncertainties of the extinction law. The distance moduli derived in each CMD are consistent with each other.
	% The solid line in the CMD is an isochrone set for the solar metallicity and log(age[yr])=8.9, 
	%We can see a few giants and subgiants around the isochrone, and 	%Stars in the inclined dash-dot box in the CMD are used for the CCD of Berkeley 49. 		

	\small
	\begin{table}[bp]
	\begin{center}
	\centering
	\caption{Estimates for reddenings and distance moduli of Berkeley 49 with different colors \label{tab:be49}}
	\doublerulesep2.0pt
	\renewcommand\arraystretch{1.5}
	\begin{tabular}{c c c}
	\hline \hline
	CMD & $E(B-V)$ & $(m-M)_0$\\
	\hline
	$r-(g-r)$ & 1.18$\pm$0.05 & 11.6$\pm$0.2\\
	$r-(g-i)$ & 1.15$\pm$0.05 & 11.6$\pm$0.2\\
	$r-(g-z)$ & 1.10$\pm$0.05 & 11.8$\pm$0.2\\
	$r-(u-g)$ & 1.05$\pm$0.05 & 11.6$\pm$0.2\\
	\hline
	\end{tabular}
	\end{center}
	\end{table}
	\normalsize

\subsection{Berkeley 84}

	The coordinates for the center of Berkeley 84 ($\alpha$ = 20h 04m 42.6s, $\delta$ = +33$\arcdeg$ 54$\arcmin$ 24$\arcsec$) derived in this study are different by -6$\arcsec$.0 in RA and 6$\arcsec$ in Dec from the values in DAML02. Figure \ref{fig:rdp}(d) shows the radial number density profile of Berkeley 84. Based on the profile, we determined the radius of Berkeley 84 to be $R_{\mathrm{cl}}=1\arcmin.25\pm0.13$.

\begin{figure}[tp]
	\centering \epsfxsize=8cm
	\epsfbox{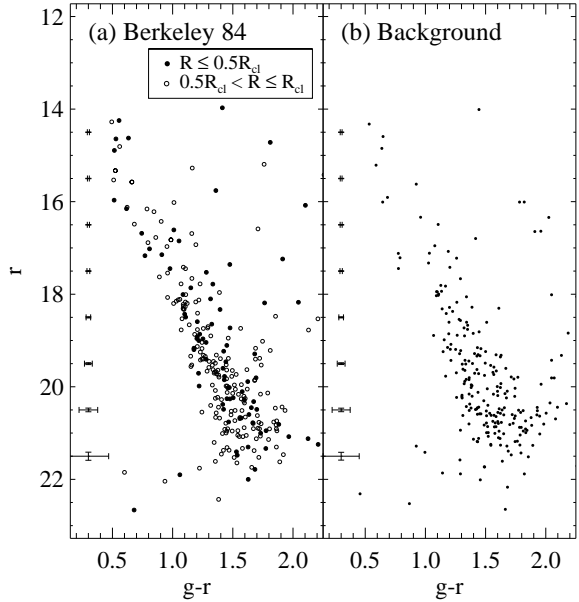}
	\caption{(a) $r-(g-r)$ CMD of the stars in Berkeley 84. Open circles are stars inside the radius of Berkeley 84, and filled dots are stars inside the half radius of the cluster. (b) $r-(g-r)$ CMD of stars in the background region, which is shown in Figure \ref{fig:image}(d). Mean errors are marked at the left side of each panel. \label{fig:be84cmd}}
	\end{figure}

	It is rather hard to study Berkeley 84 only using the SDSS because five stars inside the radius of Berkeley 84 are saturated. These bright stars ($r \leq 14$) are not shown in CMDs, and the main sequence of Berkeley 84 looks broad. Therefore the CMDs of Berkeley 84 and the background region look similar (Figure \ref{fig:be84cmd}). This may be due to severe foreground reddening. In spite of that, the number and distribution of the stars at $(g-r)\simeq0.5$ and $14\leq r \leq16$ look different from that of the background so that most of the stars in the bright main sequence are probably the cluster members.

\begin{figure}[htp]
	\centering \epsfxsize=8cm
	\epsfbox{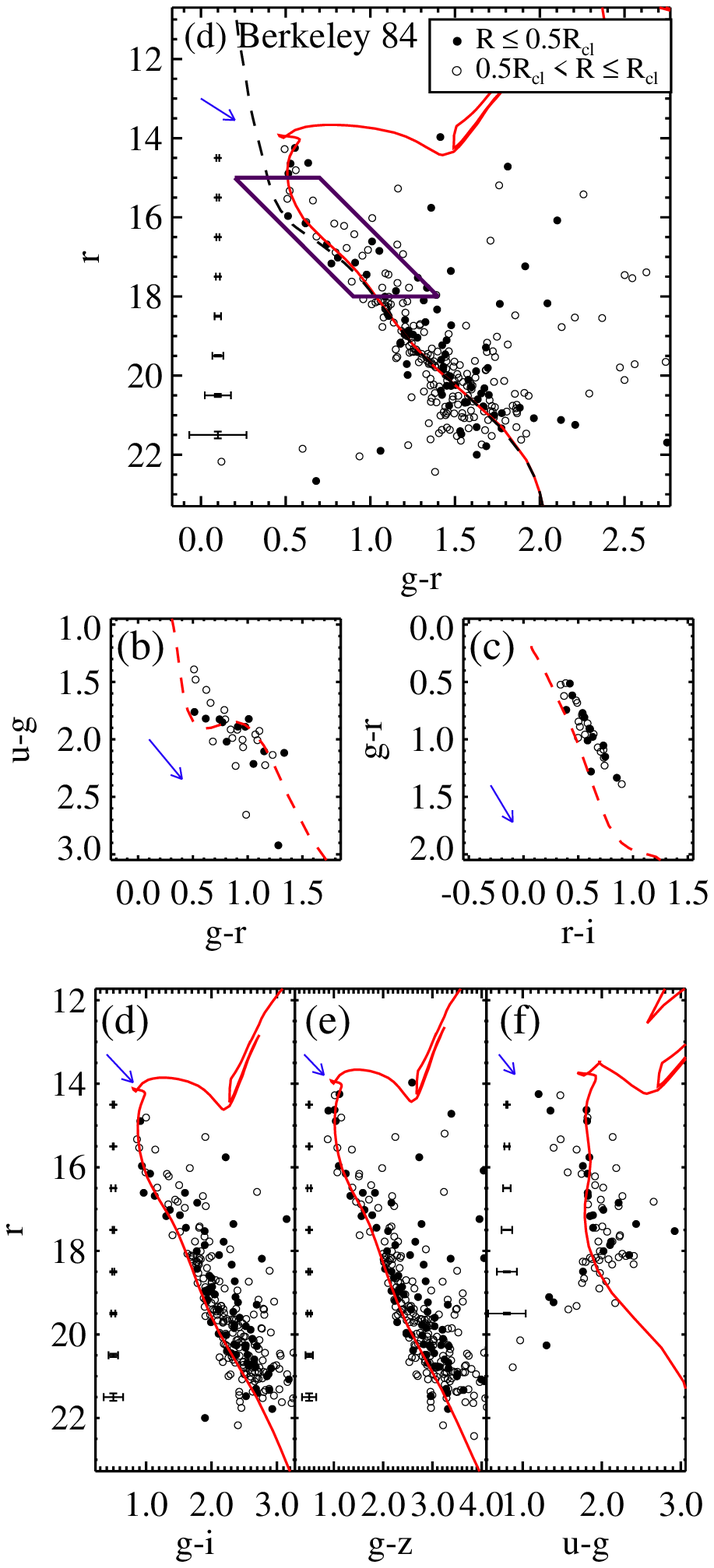}
	\caption{CMDs and CCDs of Berkeley 84. Symbols represent the same as in Figure \ref{fig:be84cmd}. The solid line represents an isochrone for log(age[yr])=8.65 and $Z=0.019$, and the dashed line represents the ZAMS. The arrow in each panel represents a reddening vector. (a) The isochrone fitting for Berkeley 84. The isochrone is shifted according to $E(B-V)=0.73$ and $(m-M)_0=12.2$. The inclined solid box contains main sequence stars which are plotted in the following CCDs. (b) $(u-g)-(g-r)$ CCD. (c) $(g-r)-(r-i)$ CCD. (d) $r-(g-i)$ CMD. (e) $r-(g-z)$ CMD. (f) $r-(u-g)$ CMD. \label{fig:be84}}
	\end{figure}

	Figure \ref{fig:be84} shows the result of the isochrone fitting. We derived its age, log(age[yr])=8.65, and the reddenings: $E(B-V)=0.73 \pm 0.06$, $0.80 \pm 0.06$, $0.73 \pm 0.06$, and $0.65 \pm 0.06$ from $r-(g-r)$ CMD, $r-(g-i)$ CMD, $r-(g-z)$ CMD, and $r-(u-g)$ CMD, respectively. Corresponding distance moduli we derived are all $(m-M)_0=12.2 \pm 0.2$. These values are also listed in Table \ref{tab:be84}. The reddening values derived from the $r-(g-r)$ and $r-(g-z)$ CMDs agree, while the reddening from the $r-(g-i)$ CMD is higher and that from the $r-(u-g)$ CMD is smaller. Figure \ref{fig:be84}(b) and (c) display the $(u-g)-(g-r)$ CCD and $(g-r)-(r-i)$ CCD for the bright main sequence stars located within the solid box in Figure \ref{fig:be84}(a). They are well-fitted by the ZAMS shifted according to the reddening $E(B-V)=0.73$.
			
	\small
	\begin{table}[bp]
	\begin{center}
	\centering
	\caption{Estimates for reddenings and distance moduli of Berkeley 84 with different colors \label{tab:be84}}
	\doublerulesep2.0pt
	\renewcommand\arraystretch{1.5}
	\begin{tabular}{c c c}
	\hline \hline
	CMD & $E(B-V)$ & $(m-M)_0$\\
	\hline
	$r-(g-r)$ & 0.73$\pm$0.06 & 12.2$\pm$0.2\\
	$r-(g-i)$ & 0.80$\pm$0.06 & 12.2$\pm$0.2\\
	$r-(g-z)$ & 0.73$\pm$0.06 & 12.2$\pm$0.2\\
	$r-(u-g)$ & 0.65$\pm$0.06 & 12.2$\pm$0.2\\
	\hline
	\end{tabular}
	\end{center}
	\end{table}
	\normalsize

\subsection{Pfleiderer 3}

	The coordinates for the center of Pfleiderer 3 ($\alpha$ = 23h 08m 11.8s, $\delta$ = +60$\arcdeg$ 51$\arcmin$ 54$\arcsec$) derived in this study are different by 12$\arcsec$.0 in RA and -30$\arcsec$ in Dec from the values in DAML02. From the radial number density profile (Figure \ref{fig:rdp}(e)), we determined the radius of Pfleiderer 3 as $R_{\mathrm{cl}}=2\arcmin.00\pm0.13$.

\begin{figure}[tp]
	\centering \epsfxsize=8cm
	\epsfbox{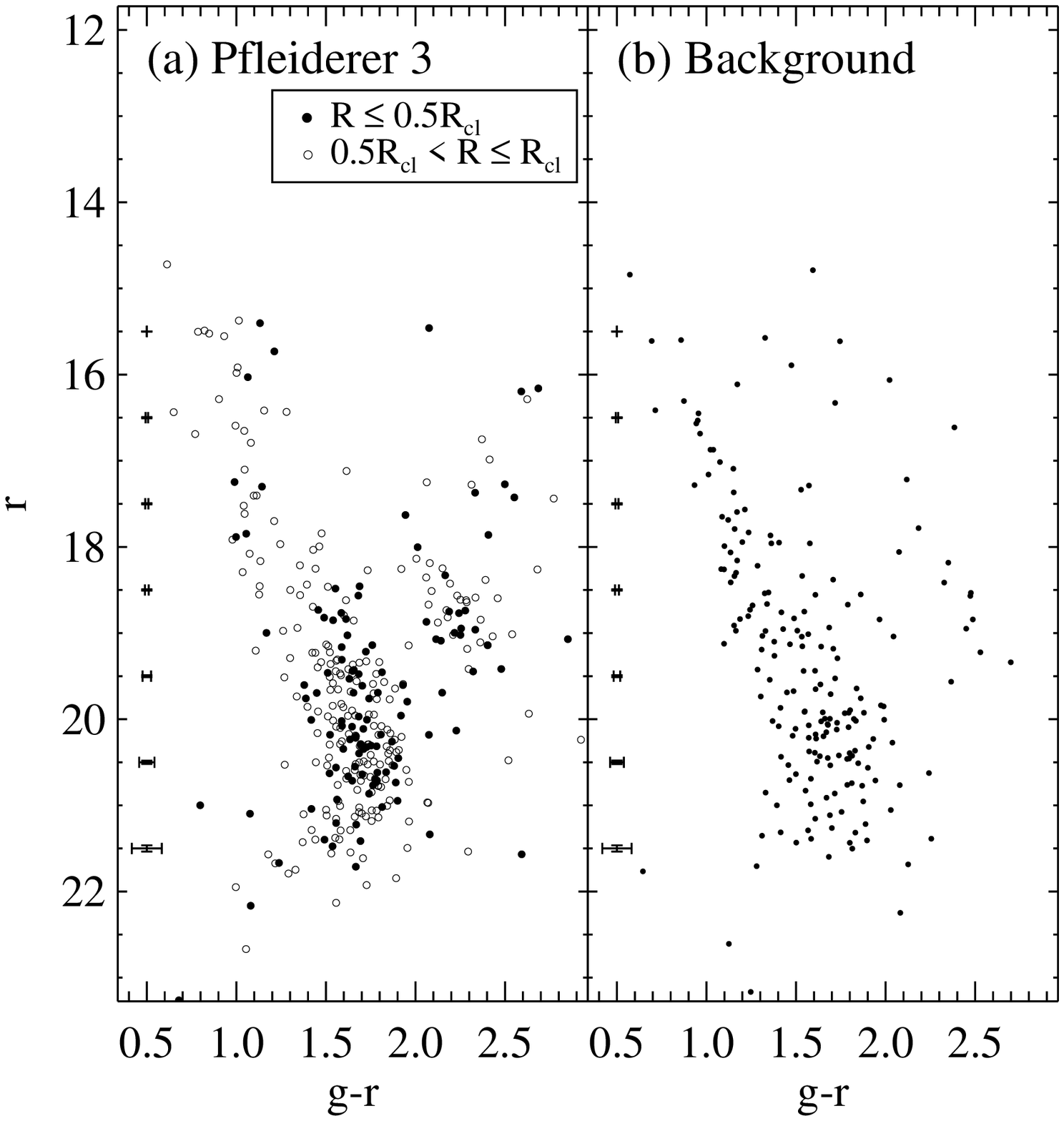}
	\caption{(a) $r-(g-r)$ CMD of the stars in Pfleiderer 3.  Open circles are stars inside the radius of Pfleiderer 3, and filled dots are stars inside the half radius of the cluster. (b) $r-(g-r)$ CMD of stars in the background region, which is shown in Figure \ref{fig:image}(e). Mean errors are marked at the left side of each panel. \label{fig:pf3cmd}}
	\end{figure}

	Pfleiderer 3 is not clearly visible in the gray-scale map of $r$ band image (Figure \ref{fig:image}(e)). However, the $r-(g-r)$ CMD of Pfleiderer 3 shows an obvious red giant clump at $(g-r)\simeq2.2$ and $r\sim19$ (Figure \ref{fig:pf3cmd}(a)). Therefore the blue and bright part of the main sequence around $(g-r)\simeq1.0$ is possibly consist of foreground stars, as shown in the CMD of the control field (Figure \ref{fig:pf3cmd}(b)).
	
\begin{figure}[tp]
	\centering \epsfxsize=8cm
	\epsfbox{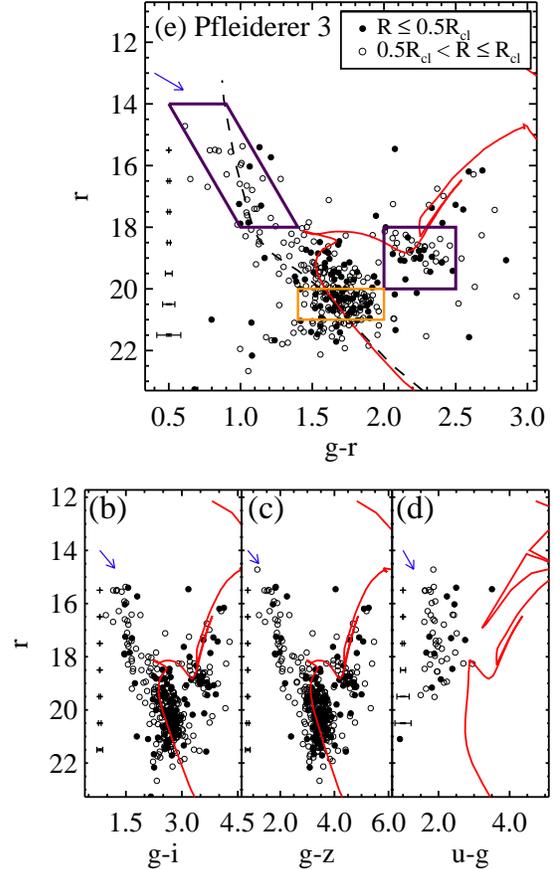}
	\caption{CMDs of Pfleiderer 3. Symbols represent the same as in Figure \ref{fig:pf3cmd}. The arrow in each panel represents a reddening vector. (a) An isochrone fitting for Pfleiderer 3. Solid line represents an isochrone for log(age[yr])=9.0 and $Z=0.016$. It is shifted according to $E(B-V)=1.5$ and $(m-M)_0=13.3$. Dashed line represents the ZAMS, that is shifted according to $E(B-V)=1.4$ and $(m-M)_0=12.95$. The inclined solid parallelogram at the left sequence contains foreground main sequence stars, and a solid square at the right contains red giant clump stars. Those stars are plotted in the following CCDs. Another square, located within $20<r<21$, contains main sequence stars. Radial distributions of stars located within each region are discussed in Section 5.4. (b) $r-(g-i)$ CMD. (c) $r-(g-z)$ CMD. (d) $r-(u-g)$ CMD. \label{fig:pf3}}
	\end{figure}

\begin{figure}[tp]
	\centering \epsfxsize=8cm
	\epsfbox{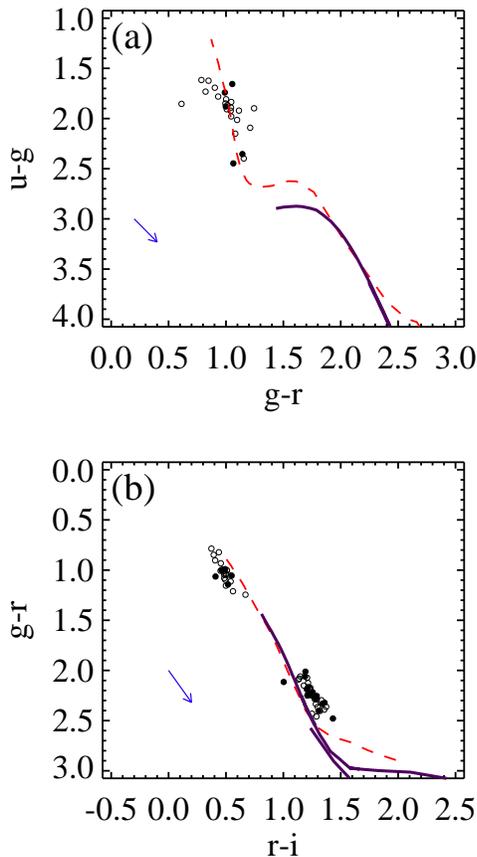}
	\caption{CCDs of Pfleiderer 3. Symbols represent the same as in Figure \ref{fig:pf3cmd}. The arrow in each panel represents a reddening vector. (a) $(u-g)-(g-r)$ CCD of the stars in Pfleiderer 3. (b) $(g-r)-(r-i)$ CCD of stars in Pfleiderer 3. The dashed line in each CCD represents the ZAMS shifted according to $E(B-V)=1.40$, and the solid line represents the giant part of an isochrone for log(age[yr])=9.0 shifted according to $E(B-V)=1.50$. \label{fig:pf3ccd}}
	\end{figure}

	Figure \ref{fig:pf3} shows the result of the isochrone fitting. We derived its age, log(age[yr])=9.0, the reddening, $E(B-V)=1.50 \pm 0.05$, and distance modulus, $(m-M)_0=13.3 \pm 0.2$ from the $r-(g-r)$ CMD. Values derived from all other CMDs agree with those values. We also derived the reddening and distance modulus for the foreground sequence using the ZAMS. The reddening is $E(B-V)=1.40 \pm 0.05$, and the distance modulus is $(m-M)_0=12.95 \pm 0.2$. Figure \ref{fig:pf3ccd} displays the $(u-g)-(g-r)$ CCD and $(g-r)-(r-i)$ CCD for the bright foreground sequence stars and the red giant clump stars located within two boxes in Figure \ref{fig:pf3}(a). The red giant clump stars are too faint at the $u$ band, so that they do not appear in the $(u-g)-(g-r)$ CCD (Figure \ref{fig:pf3ccd}(a)), neither $r-(u-g)$ CMD (Figure \ref{fig:pf3}(d)). Figure \ref{fig:pf3ccd}(b) shows the $(g-r)-(r-i)$ CCD. There are two groups of stars: the blue one is the foreground sequence stars and the red one is the red giant clump stars.
	
%%%%%%%%%%%%%%%%%%%%%%%%%%%%%%%%%%%%%%%%%%%%%%%%%%%%%%%%%%%%%%%%%%%%%%%%%%%%%%%%%%%
%%%%%%%%%%%%%%%%%%%%%%%%%%%%%%%%%%%%%%%%%%%%%%%%%%%%%%%%%%%%%%%%%%%%%%%%%%%%%%%%%%%	
\section{DISCUSSION}
\subsection{Comparison with previous studies}

	The parameters of the clusters derived in this study are summarized in Table \ref{tab:param}. Three clusters (Czernik 5, Berkeley 49, and Berkeley 84) among our targets were studied previously \citep{ta08,ta09,cam09,su10}. We compare these for three clusters with those in the previous studies.

	In the case of Czernik 5, \citet{ta09} determined its age to be 700 Myrs (log(age[yr])=8.85), considering the existence of some red giant branch stars in the CMD based on 2MASS photometry. This value is much larger than our estimate, $280\pm100$ Myr. This large difference is probably due to the fact that we used the cluster coordinates derived in this study, which are different from those used by \citet{ta09}, as $560\arcsec$ in RA and $81\arcsec$ in Dec.

	The physical parameters of Berkeley 49 were derived by \citet{ta08} from 2MASS photometry, and by \citet{su10} from $BVI$ photometry.	We obtain the value for the age of Berkeley 49, $794\pm210$ Myrs (log(age[yr])=$8.9\pm0.13$), while \citet{ta08} derived 160 Myrs (log(age[yr])=8.2) and \citet{su10} obtained $270\pm46$ Myrs (log(age[yr])=$8.43\pm0.07$). Both of these previous studies estimated that the age of Berkeley 49 is younger than our study.
	
\begin{figure}[tp]
	\centering \epsfxsize=8cm
	\epsfbox{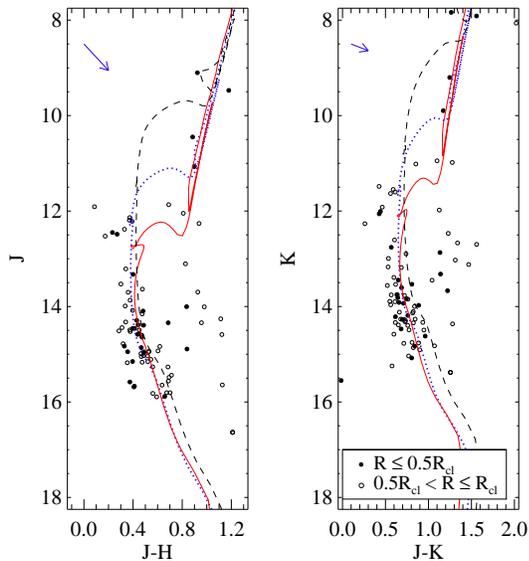}
	\caption{\textit{Left}. $J-(J-H)$ CMD of the stars in Berkeley 49. \textit{Right}. $K-(J-K)$ CMD. Solid line represents the isochrone for the same parameters as determined in the $r-(g-r)$ CMD (Figure \ref{fig:be49}(a)). Dashed line and dotted line represent the isochrones for the parameters given in \citet{ta08} and \citet{su10}, respectively. \label{fig:be49cmd_2mass}}
	\end{figure}

	The age difference between ours and \citet{ta08} may be due to the fact that we included RGB stars as the members, while \citet{ta08} did not. These stars, shown in Figure \ref{fig:be49cmd_2mass}, have similar proper motions to those for other members of Berkeley 49 so that they are considered to be the members of this cluster. It is noted that the four brightest RGB stars ($J\leq11.2$) are located within the half radius of Berkeley 49. Therefore we consider that these RGB stars are members of the cluster, while \citet{ta08} regarded these stars as field stars. 

	The age difference between ours and \citet{su10} may be due to a difference in the turnoff point in the CMD. We considered the stars at $r\simeq16.0$ mag and $(g-r)\simeq1.1$ to be a turnoff point, and considered nine brighter stars above this turnoff to be field stars. In contrast, \citet{su10} considered the turnoff point to be at $V\simeq15$ mag and $(B-V)\simeq1.2$, (corresponding to $r\simeq14.6$ mag and $(g-r)\simeq1.0$ \citep{jes05}), which is 1.4 mag brighter than the turnoff we chose.

	The difference in age estimates may depend on the reddening difference. Our reddening estimate, $E(B-V)=1.18$, is slightly smaller than the values given by \citet{ta08} and \citet{su10}, $E(B-V)=1.57$ and $E(B-V)=1.35$, respectively. Therefore the larger values	for ages are derived when we choose smaller values for reddenings.
	
	In spite of these reddening and age differences, estimated distances are marginally consistent with each other within the error. Our distance is $2090\pm200$ pc ($(m-M)_0=11.6\pm0.2$), while the distance of \citet{ta08} is $2035\pm110$ pc ($(m-M)_0=11.54\pm0.12$) and that of \citet{su10} is $2300\pm230$ pc ($(m-M)_0=11.81\pm0.22$).
				
\begin{figure}[tp]
	\centering \epsfxsize=8cm
	\epsfbox{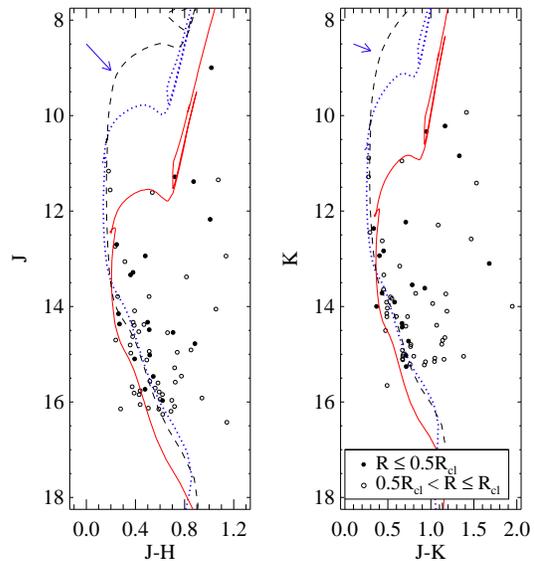}
	\caption{The same as Figure \ref{fig:be49cmd_2mass} but for Bekeley 84. Dashed line and dotted line represent the isochrones for the parameters given in \citet{ta08} and \citet{cam09},	respectively. \label{fig:be84cmd_2mass}}
	\end{figure}

	We also compared the physical parameters of Berkeley 84 with those in \citet{ta08} and \citet{cam09}, using the 2MASS CMDs which are shown in Figure \ref{fig:be84cmd_2mass}. Both of the previous studies regarded four RGB stars around $J\simeq11.5$ mag and $(J-H)\simeq0.8$ as field stars, and considered that two blue stars around $J\simeq11.5$ mag and $(J-H)\simeq0.2$ to be the members. However, the brightest RGB star with $J\simeq9$ mag and two RGB stars with $J\simeq11.5$ mag are located within the half radius of Berkeley 84, while the two blue stars are located at the distance farther the half radius. Therefore the RGB stars are considered to have a higher probability of being the members than the blue stars. This different member selection is a reason for the different estimates of reddenings. Our reddening value for $E(B-V)=0.73\pm0.06$ is consistent with the value of \citet{ta08}, $E(B-V)=0.76\pm0.1$. Our estimate is much larger than the value derived by \citet{cam09}, $E(B-V)=0.58\pm0.06$. 
				
	The reason for age differences of Berkeley 84 seems to be similar to the case for Berkeley 49. We derived the age of Berkeley 84, $447\pm130$ Myrs (log(age[yr])=$8.65\pm0.15$). \citet{ta08} and \citet{cam09} estimated the age to be 120 Myrs (log(age[yr])=8.1) and $360\pm50$ Myrs (log(age[yr])=$8.55\pm0.06$), respectively. \citet{ta08}'s value is much smaller than ours, while \citet{cam09}'s value is consistent with ours within the error. Our distance estimate, $2750\pm270$ pc ($(m-M)_0=12.2\pm0.2$), is larger than $2025\pm95$ pc ($(m-M)_0=11.55\pm0.12$) for \citet{ta08} and $1700\pm100$ pc ($(m-M)_0=11.15\pm0.13$) for \citet{cam09}.

	\small
	\begin{deluxetable}{c c c c c c c}
	\tablecolumns{8}
	\tablewidth{0pc}
	\tablecaption{Physical parameters of the clusters \label{tab:param}}
	\tablehead{
		\colhead{Name} &
		\colhead{Age} &
		\colhead{$E(B-V)$} &
		\colhead{Distance} &
		\colhead{$R_{\mathrm{GC}}$\tablenotemark{a}} &
		\colhead{z\tablenotemark{b}} &
		\colhead{Reference}\\
		\colhead{} &
		\colhead{Myr} &
		\colhead{mag} &
		\colhead{pc} &
		\colhead{kpc} &
		\colhead{pc}
	}
	\startdata
		Czernik 5 & 280 $\pm$ 100 & 1.20 $\pm$ 0.05 & 2750 $\pm$ 270 & 10.01 $\pm$ 0.20 & $-27 \pm 3$ & This study\\
		& 700 & 1.23 & 2205 $\pm$ 100 & 9.57 $\pm$ 0.07 & $-23$\tablenotemark{c} & \citet{ta09}\\
		Alessi 53 & 250 $\pm$ 150 & 0.80 $\pm$ 0.05 & 4570 $\pm$ 330 & 12.35 $\pm$ 0.67 & $-53 \pm 4$ & This study\\
		Berkeley 49 & 794 $\pm$ 210 & 1.18 $\pm$ 0.05 & 2090 $\pm$ 200 & 7.58 $\pm$ 0.01 & $\;\;\:94 \pm 9$ & This study\\
		& 160 & 1.57 $\pm$ 0.1 & 2035 $\pm$ 110 & 7.58 $\pm$ 0.01 & $\;\;\:91$\tablenotemark{c} & \citet{ta08}\\
		& 270 $\pm$ 46 & 1.35 $\pm$ 0.02 & 2300 $\pm$ 230 & 7.57 $\pm$ 0.01 & $\;\;\:19 \pm 2$ & \citet{su10}\\
		Berkeley 84 & 447 $\pm$ 130 & 0.73 $\pm$ 0.06 & 2750 $\pm$ 270 & 7.57 $\pm$ 0.01 & $\;\;\:61 \pm 6$ & This study\\
		& 120 & 0.76 $\pm$ 0.1 & 2025 $\pm$ 95 & 7.58 $\pm$ 0.01 & $\;\;\:45$\tablenotemark{c} & \citet{ta08}\\
		& 360 $\pm$ 50 & 0.58 $\pm$ 0.06 & 1700 $\pm$ 100 & 7.61 $\pm$ 0.01 & $\;\;\:38 \pm 3$ & \citet{cam09}\\
		Pfleiderer 3 & 1000 $\pm$ 260 & 1.50 $\pm$ 0.05 & 4570 $\pm$ 440 & 10.53 $\pm$ 0.31 & $\;\;\:37 \pm 4$ & This study\\
	\enddata
	\tablenotetext{a}{All $R_{\mathrm{GC}}$ are calculated using $R_\odot=8$ kpc.}
	\tablenotetext{b}{These values represent heights from the galactic plane. All these values are calculated using their distances, except \citet{ta08} and \citet{ta09}.}
	\tablenotetext{c}{These values are presented in the references.}
	\end{deluxetable}
	\normalsize
		
\subsection{Spatial distribution of the open clusters}

\begin{figure}[tp]
	\centering \epsfxsize=8cm
	\epsfbox{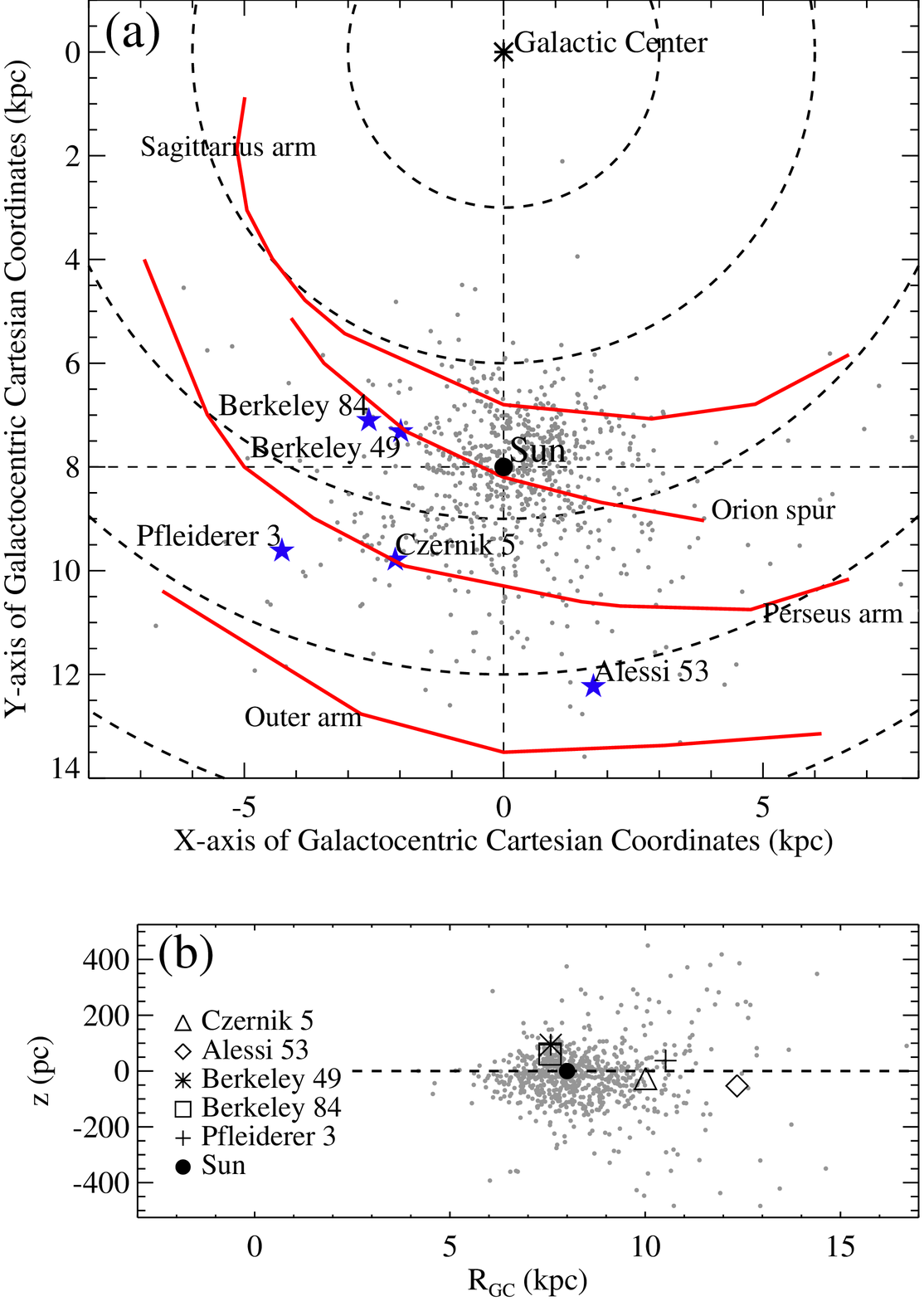}
	\caption{(a) The spatial distribution of the clusters (star symbols) in the Galactic plane. Dots represent the intermediate-age clusters ($7.4<$ log(age[yr]) $<9.0$) in DAML02. The Sun is marked by a filled circle at the center. The radii of the dashed line circles are 3, 6, 9, 12, and 15 kpc. Curved solid lines represent spiral arms. (b) The vertical distribution of the clusters with respect to the Galactic plane (dashed line). Dots represent the same as in (a). Large symbols represent the target clusters and the Sun, respectively. \label{fig:spd}}
	\end{figure}

	With the distance estimates for our target clusters we can investigate where they are located in the Galactic plane. In Figure \ref{fig:spd}(a) we show the spatial distribution of these clusters in the galactic plane. We also plotted the positions of intermediate-age clusters in DAML02 for comparison. The age distribution of the clusters in DAML02 shows three groups: young clusters with log(age[yr])$<7.4$, intermediate-age clusters with	log(age[yr])$=7.4-9.0$, and old clusters with log(age[yr])$>9.0$. The age range for the intermediate-age clusters covers the ages of our target clusters. Therefore we plotted the intermediate-age clusters with  log(age[yr])$=7.4-9.0$. In addition we marked the spiral arms in our Galaxy \citep{chu09}. The intermediate-age clusters are mostly located near the Sun, and they do not show any significant correlation with the arms. Berkeley 49 and Berkeley 84 are located in the Orion spur, Czernik 5 is in the Perseus arm, and Pfleiderer 3 and Alessi 53 are located between the Perseus arm and the outer arm.

	Figure \ref{fig:spd}(b) displays the distance from the Galactic plane ($z$) versus the galactocentric distance ($R_{\mathrm{GC}}$) for the target clusters. We also plotted the same for the intermediate-age clusters in DAML02 for comparison. All the target clusters are closer than 94 pc to the Galactic plane.	 

\subsection{The blue stars in Berkeley 49}
	
	We found nine blue stars in Berkeley 49. These stars are bluer than the turnoff point of Berkeley 49, and redder than the ZAMS. \citet{su10} considered that these stars are the most bright main sequence stars of Berkeley 49. However, they seem to be neither the main sequence stars nor blue straggler stars, are described in the following.
	
	It is considered that blue straggler stars are heavier than main sequence stars. Therefore if these blue stars are blue straggler stars or main sequence stars, then it is expected that they show a stronger or similar central concentration than the other main sequence stars. We investigated this using two ways: the spatial distribution of the stars, and the cumulative distribution functions (CDFs) for radial distributions of the stars. We selected three groups of stars in this cluster for the analysis: the blue stars, the bright main sequence stars, and the faint main sequence stars, according to the marked regions in Figure \ref{fig:be49}(a).

\begin{figure}[tp]
	\centering \epsfxsize=8cm
	\epsfbox{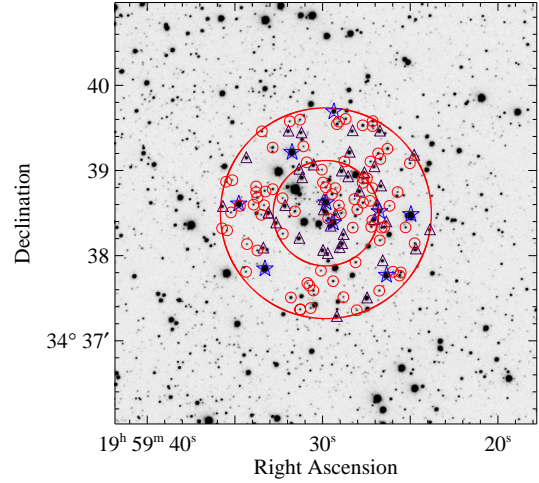}
	\caption{A gray-scale map of Berkeley 49. The large solid circle represents the radius of the cluster, and the small circle represents the half radius. Stars, triangles, and open circles represent the blue stars, bright main sequence stars, and faint main sequence stars, respectively. \label{fig:be49map}}
	\end{figure}

	Figure \ref{fig:be49map} shows the spatial distribution of these three groups of stars marked in a gray-scale map of Berkeley 49. The bright main sequence stars show a stronger central concentration than the other two groups. It is also seen that both the blue stars and the faint main sequence stars are distributed sparsely all over the cluster region.

\begin{figure}[tp]
	\centering \epsfxsize=8cm
	\epsfbox{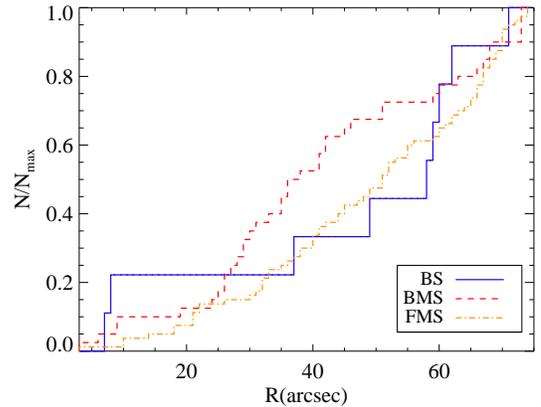}
	\caption{Cumulative number distributions of stars in Berkeley 49. The solid, dashed, and dash-dot line represent the blue stars, bright main sequence stars, and faint main sequence stars, respectively. Numbers of the stars are normalized by the maximum value.\label{fig:be49_cdf}}
	\end{figure}

	In Figure \ref{fig:be49_cdf} we plotted the CDFs for radial distributions of the three groups of stars. It is seen that the bright main sequence stars are more centrally concentrated than the other two groups, and that these two groups show similar distributions. We performed Kolmogorov-Smirnov (K-S) tests to estimate how much these groups are different in radial distribution. The tests yield 19\%, 72\%, and 7\% probabilities for the sets of (blue stars -- bright main sequence stars), (blue stars -- faint main sequence stars), and (bright -- faint main sequence stars), respectively. These probabilities indicate that the two main sequence groups are clearly different. However it is not obvious that the blue stars are different with the other two groups. This is not consistent with the dynamic prediction for the blue straggler stars. Therefore the blue stars are considered to be field stars.

\subsection{The blue sequence of Pfleiderer 3}
			
	We found a blue sequence in the CMD of Pfleiderer 3 (Figure \ref{fig:pf3}(a)). We selected the blue sequence stars, the red giant clump stars, and the main sequence stars in this cluster to check the membership of the former, according to the marked regions in Figure \ref{fig:pf3}(a).

\begin{figure}[tp]
	\centering \epsfxsize=8cm
	\epsfbox{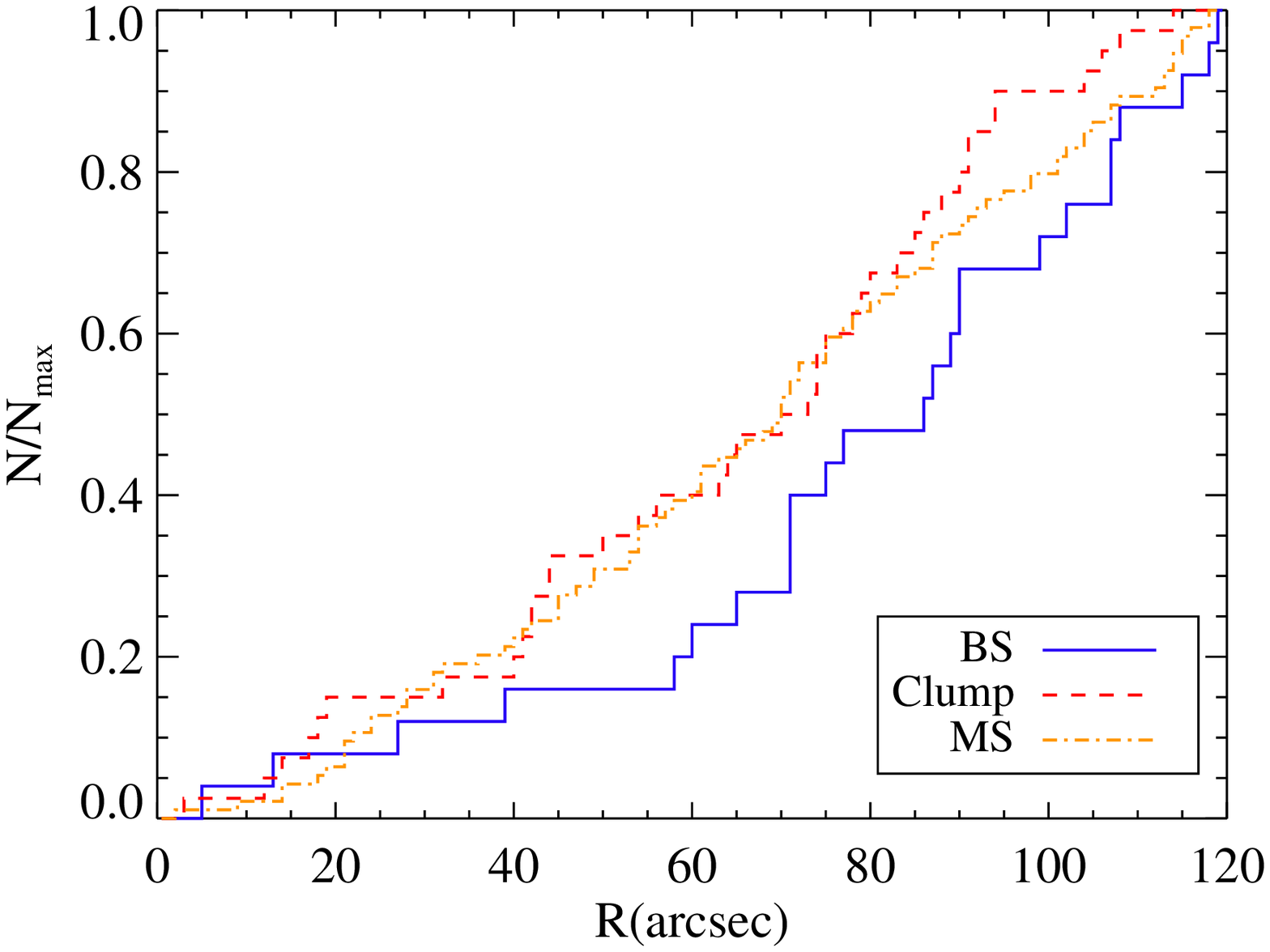}
	\caption{Cumulative number distributions of stars in Pfleiderer 3. The solid and dashed line represent the blue sequence stars and the red giant clump stars, respectively. Numbers of the stars are normalized by the maximum value.\label{fig:pf3_cdf}}
	\end{figure}

	Figure \ref{fig:pf3_cdf} shows CDFs for radial distributions of the blue sequence stars, the red giant clump stars, and the main sequence stars in Pfleiderer 3. The red giant clump stars and the main sequence stars show a stronger central concentration than the blue sequence stars. Therefore these blue sequence stars are considered to be foreground stars. K-S tests yield 27\% probability for the set of (blue sequence stars -- red giant clump stars) and 66\% probability for the set of (blue sequence stars -- main sequence stars). These probability indicate that the blue sequence stars are different population from the red giant clump stars and the main sequence stars.
	
	According to the estimates of the distance and foreground reddening of this blue sequence, the blue sequence stars, being 680 pc closer than Pfleiderer 3, are located in the Perseus arm, while Pfleiderer 3 is between the Perseus arm and the outer arm. It is consistent with the location of Pfleiderer 3 which is at beyond the Perseus arm. 

%%%%%%%%%%%%%%%%%%%%%%%%%%%%%%%%%%%%%%%%%%%%%%%%%%%%%%%%%%%%%%%%%%%%%%%%%%%%%%%%%%%
%%%%%%%%%%%%%%%%%%%%%%%%%%%%%%%%%%%%%%%%%%%%%%%%%%%%%%%%%%%%%%%%%%%%%%%%%%%%%%%%%%%	
\section{SUMMARY}
	We have derived the physical parameters of five open clusters in the SDSS. To obtain parameters, we determine the position of these clusters first. Then we exclude fast proper motion stars using the velocity dispersion of each cluster using the PPMXL. The size of the clusters is determined using remaining stars. We assume the metallicity of these clusters to follow the radial metallicity gradient of OCs using data in DAML02. We determine the age, distance, and reddening of the clusters using the Padova isochrones. The ages of the clusters are in the range from 250 to 1000 Myrs, the distance to these clusters are derived to be $2.0-4.4$ kpc. The reddenings are estimated to be $E(B-V)=0.71-1.55$. The derived physical parameters are listed in Table \ref{tab:param}.
	
	These clusters follow the distribution of the intermediate-age OCs without exception. They are located from the Orion spur to between the Perseus arm and the outer arm. The distance from the galactic plane for the clusters is less than 94 pc.
	
	We find nine stars which are bluer than the turnoff point of Berkeley 49. We compare their central concentration with that of bright and faint main sequence stars. The blue stars are not more centrally concentrated than the bright main sequence stars, but are distributed sparsely all over the cluster region. Therefore these blue stars are considered to be field stars.
	
	There are many blue sequence stars in Pfleiderer 3, but they seem to be foreground stars. Physical parameters of these blue sequence stars derived from the ZAMS fitting. The reddening is derived to be $E(B-V)=1.40\pm0.05$ and distance modulus is $(m-M)_0=12.95\pm0.2$. This is corresponding to the distance to the Perseus arm lying the foreground of the cluster.

%--------------------------------------------------------------------
\acknowledgments{
This work was supported by Mid-career Researcher Program through NRF grant funded by the MEST (No.2010-0013875). This research was supported also by the BK21 program of the Korean Government.}

%--------------------------------------------------------------------

%-------------------------------------------------------------------
\end{document}